\documentclass[11pt,a4paper]{article}
\usepackage{jheppub}
\usepackage{subcaption}
\usepackage{slashed}
\usepackage[normalem]{ulem}
\usepackage{bbold}
\usepackage{tikz}
\usepackage{tikz-feynman}
\usepackage{contour}
\usepackage{xcolor}
\usepackage{tikz}
\usepackage{hyperref}

\definecolor{lime}{HTML}{A6CE39}
\DeclareRobustCommand{\orcidicon}{%
	\begin{tikzpicture}[baseline=-0.6ex]
	\draw[lime, fill=lime] (0,0) circle [radius=0.16]
	node[white] {\tiny\sffamily ID};
	\draw[white, fill=white] (-0.07,0.1) circle [radius=0.01];
	\end{tikzpicture}%
}

\newcommand{\orcidlink}[1]{\href{https://orcid.org/#1}{\orcidicon}}

\makeatletter
\gdef\@fpheader{}
\makeatother
\definecolor{lime}{HTML}{A6CE39}
\DeclareRobustCommand{\orcidicon}{\hspace{-4pt}
	\begin{tikzpicture}
	\draw[lime, fill=lime] (0,0) 
	circle [radius=0.16] 
	node[white] {\hspace{0.1mm}{\fontfamily{qag}\selectfont \tiny ID}};
	\draw[white, fill=white] (-0.07,0.1) 
	circle [radius=0.01];
	\end{tikzpicture}
	\hspace{-3.2mm}
}
\definecolor{darkgreen}{rgb}{0,0.5,0}

\newcommand{\lmultau}{L_{\mu}-L_{\tau}}
\newcommand{\op}{\mathcal{O}}

\title{Muonphilic asymmetric dark matter at a future muon collider}

\author[a]{Arnab Roy\,\orcidlink{0000-0003-0249-1440}}
\author[b]{Raymond R. Volkas\,\orcidlink{0000-0002-4254-8520}}

\affiliation[a]{%\small
	School of Physics and Astronomy, Monash University, Wellington Road, Clayton, Victoria 3800, Australia}
\affiliation[b]{%\small 
	ARC Centre of Excellence for Dark Matter Particle Physics, School of Physics, The University of Melbourne, Victoria 3010, Australia}

\emailAdd{arnab.roy1@monash.edu}
\emailAdd{raymondv@unimelb.edu.au}

\abstract{We explore phenomenological constraints on, and future muon collider sensitivities to, the parameter spaces of various muonphilic portals to fermionic asymmetric dark matter (ADM). Both WEFT-level dimension-6 effective operators and two UV models based on gauged $L_\mu - L_\tau$ are considered. One of the latter features a vector coupling to the dark matter and the other an axial vector coupling. The ADM criterion that at least $99\%$ of the dark matter relic density is asymmetric is also imposed. We identify which of these scenarios are currently allowed by direct detection and collider constraints, and then determine how much more of the parameter space could be probed by 3 and 10 TeV muon colliders with 1 ab$^{-1}$ of data. For the UV models, the constraints from $g-2$ of the muon are included. The future sensitivity curves due to neutron star heating considerations are also depicted. We present results for both the few-GeV dark matter mass regime motivated by ADM approaches to the $\Omega_b \simeq \Omega_\text{DM}/5$ coincidence problem, and for larger masses in the context of more general ADM.
}
\arxivnumber{}

\begin{document}
	
	\maketitle
	
	\newpage

	\section{Introduction}
	Lepton-portal dark matter (DM), where the dominant non-gravitational couplings are to Standard Model (SM) leptons rather than quarks, provides a well-motivated scenario for DM~\cite{Baltz:2002we,Chen:2008dh,Fox:2008kb,Cao:2009yy,Cohen:2009fz,Ibarra:2009bm,Bi:2009uj,Kopp:2009et,Ko:2010at,Chao:2010mp,Schmidt:2012yg,Das:2013jca,Schwaller:2013hqa,Bai:2014osa,Agrawal:2014ufa,Kopp:2014tsa}. A particularly interesting subset is \emph{muonphilic} DM, which arises naturally in gauge extensions such as $U(1)_{L_\mu-L_\tau}$~\cite{Foot:1990mn,He:1990pn,He:1991qd,Foot:1994vd}. In such scenarios, DM couples predominantly to second-generation leptons via a new mediator, while tree-level interactions with quarks are absent. Nuclear recoils then appear only through loop-induced couplings to the electromagnetic current, bringing in suppressions from velocity, momentum transfer and the muon mass~\cite{Kopp:2009et}, so that even the latest nucleon-based direct-detection searches may still leave a certain muonphilic parameter space essentially unexplored.
	
	On another track, the observed near coincidence between the baryon and DM energy densities, $\Omega_{\rm b}\sim \Omega_{\rm DM}/5$, suggests that the two sectors might share a common origin, with comparable number densities set by related asymmetries. This is the basic idea of \textit{asymmetric dark matter (ADM)}: a primordial asymmetry between DM and anti-DM is generated in the early universe, and after interactions with the SM freeze out, the relic abundance is primarily determined by this asymmetry rather than by the freeze out of a symmetric population (see Refs.~\cite{Petraki:2013wwa,Zurek:2013wia} for comprehensive reviews and citations to the original literature).\footnote{Note that using ADM to address the coincidence problem favours a DM mass in the few-GeV regime. However, non-self-conjugate DM is a logical possibility independent of the coincidence problem. The DM asymmetry may then be quite different from the baryon asymmetry and so in this paper we consider a wide range for the DM mass scale.} Lepton-portal realisations are especially natural in this context: if the dark asymmetry is tied to the lepton sector (for instance through shared chiral gauge interactions or sphaleron-like processes), muonphilic portals offer a minimal way to communicate the asymmetry to the dark sector without introducing large couplings to quarks~\cite{Cohen:2009fz,Blennow:2010qp,Falkowski:2011xh}. 
	
	Furthermore, these kinds of portals assist with a key requirement in ADM cosmologies: the annihilation of the symmetric part of the dark matter primordial soup into radiation. Given the relatively stringent upper bound from cosmic microwave background observations on dark radiation, these annihilations must primarily produce SM particles, while adhering to various phenomenological constraints~\cite{Roy:2024ear}. These constraints are less severe for DM interactions with the higher-generation leptons.
	
	Muonphilic DM is subject to a variety of complementary constraints, such as loop-induced scattering off nuclei that is probed in large direct-detection experiments, DM--lepton scattering in compact stars that can anomalously heat old neutron stars, and LHC searches for light $Z'$ bosons in $Z\to 4\mu$ final states. In this context, a high-energy muon collider may offer a qualitatively different way to explore any remaining allowed space -- an important focus of the present study. A muon collider combines the clean environment of a lepton machine with the energy reach usually associated with hadron colliders~\cite{Delahaye:2019omf,Long:2020wfp,MuonCollider:2022xlm,Accettura:2023ked,InternationalMuonCollider:2024jyv}. It is uniquely suited to directly testing new physics that couples preferentially to muons. In particular, dark sectors interacting with muons can be probed through missing-energy signatures such as mono-photon events,
	\begin{align}
	\mu^+ \mu^- \to \chi \bar{\chi} \gamma,
	\end{align}
	where the photon arises from initial-state radiation and the DM pair escapes undetected. This process is directly sensitive to the $\mu$–DM interaction, avoids hadronic uncertainties, and can reach energy scales and mediator masses that are challenging to access at hadron colliders or in fixed-target experiments.
	
	DM and muonphilic interactions at a muon collider have already been explored from several angles. Studies of electroweak WIMP DM show that a high-energy muon collider can probe multi-TeV masses through inclusive missing-energy and disappearing-track signatures~\cite{Han:2020uak,Capdevilla:2024bwt}, while mono-photon spectra have been proposed as a general probe of dark photons, ALPs, and other light invisible particles or nonstandard neutrino interactions at muon colliders~\cite{Casarsa:2021rud,Jana:2023ogd,Chen:2025ewx}. More recently, lepton-portal and lepton-flavoured DM models have been analysed at muon colliders, focusing on symmetric freeze-out or freeze-in regimes and on the resulting prompt or long-lived signatures of the mediators~\cite{Jueid:2023zxx,Asadi:2023csb,Asadi:2024jiy,Liu:2022byu}. In parallel, gauged $U(1)_{\lmultau}$ and more general leptophilic $Z'$ scenarios have been studied at muon and other lepton colliders, including mono-photon channels, primarily to constrain the new gauge boson and its connection to the muon $g-2$ and lepton-flavour anomalies rather than to DM~\cite{Feng:2012jn,Huang:2021nkl,Dasgupta:2023zrh,De:2025hay}. What is still lacking, and what we undertake here, is a systematic treatment of \emph{muonphilic asymmetric dark matter}.
	
	At the effective field theory (EFT) level, we enumerate all independent dimension-6 four-fermion operators coupling a Dirac DM candidate $\chi$ to the muon current and classify their annihilation properties in terms of the leading partial wave. We then specialise to two minimal $U(1)_{L_\mu-L_\tau}$ UV completions that realise particularly relevant EFT structures: a vector model with a Dirac $\chi$ carrying vector-like $L_\mu-L_\tau$ charge, and an axial model in which mixing between two dark fermions yields a predominantly axial $Z'$ coupling for the lightest state. For each case, we identify the regions of parameter space where the ADM requirement -- that the relic DM density is driven mainly by the asymmetry -- can be satisfied while remaining consistent with existing astrophysical, collider and muon $g-2$ constraints (as relevant), and we assess how much of this region can be probed through mono-photon searches at a future muon collider.
	
	The paper is organised as follows. In Sec.~\ref{sec:theory} we introduce the EFT framework and the two $U(1)_{L_\mu-L_\tau}$ UV completions, and summarise their annihilation rates. This is followed in Sec.~\ref{sec:relicdensity} by a discussion of the relic density of ADM. Section~\ref{sec:muoncollider} then describes the muon collider setup and the mono-photon analysis strategy. In the subsequent sections \ref{sec:other_bounds} and \ref{sec:global_bounds} we collect the relevant direct detection, muon $g-2$, collider and trident bounds, combine them, as relevant, with the ADM requirement, and then compare them with the projected muon collider reach for both EFT and UV scenarios and prospective neutron star heating sensitivities for the EFT cases. We conclude with a summary.

	\section{Theoretical framework}
	\label{sec:theory}
	
	We consider a Dirac fermion DM candidate, $\chi$, motivated by ADM, whose dominant non-gravitational interactions are with muons of the SM via a new mediator that acts as a portal between the dark and visible sectors. The appropriate theoretical description of the scenario depends on the mediator mass, $m_{\rm med}$, relative to $m_\chi$. For a heavy mediator, $m_{\rm med}\gg m_\chi$, the dynamics are captured by an EFT involving only SM and DM degrees of freedom~\cite{Beltran:2010ww,Goodman:2010ku}. For a light mediator, $m_{\rm med} \lesssim \mathcal{O}(m_\chi)$, the EFT breaks down and the use of an ultraviolet (UV) completion is required. In what follows, we list the relevant muonphilic EFT operators for the former case and provide representative UV completions corresponding to two of the EFT operators.

	\subsection{EFT contact-interactions}
	
	In this case, below a cut-off scale $\Lambda$, we only have one BSM particle $\chi$ with all other BSM particles having decoupled. We can then expand the SM lagrangian by higher dimensional operators $\mathcal{O}^{(n)}_i$ with dimension $n$, and Wilson coefficients $\mathcal{C}_i$,
	\begin{align}
	\mathcal{L}=\mathcal{L}_{SM}+\sum_i \frac{\mathcal{C}_{i}^{(5)}}{\Lambda} \mathcal{O}^{(5)}_i+\sum_i \frac{\mathcal{C}_{i}^{(6)}}{\Lambda^2} \mathcal{O}^{(6)}_i + \ldots
	\end{align}
	Note that $\Lambda$ does not set a strict cut-off for the EFT --- the actual breakdown scale depends on the mediator masses and couplings~\cite{Busoni:2013lha,Contino:2016jqw}. Assuming perturbative s-channel UV mediation, as in our case, a conservative EFT validity condition is $\Lambda\gtrsim m_\chi/2\pi$~\cite{Busoni:2013lha,Shoemaker:2011vi}. 
	
	The relevant dimension-5 interactions are Higgs portal couplings, which are strongly constrained by invisible Higgs decay searches~\cite{CMS:2022qva,ATLAS:2022yvh}. For example, for fermionic DM with $m_\chi \lesssim 20\,$GeV, Higgs portal interactions consistent with $\mathrm{BR}(H_{\rm SM} \to \mathrm{inv})<10\%$ predict spin-independent scattering below the neutrino floor~\cite{Strigari:2009bq,Billard:2013qya,Monroe:2007xp}, making them irrelevant for direct detection~\cite{Arcadi:2021mag}. For larger masses, the relevant parameter space is excluded by current limits~\cite{Arcadi:2021mag}.
	
	Restricting to dimension-6 contact interactions with muons only, the relevant operators take the generic form
	\begin{align}
	\mathcal{O}_{\Gamma\Gamma'} = 
	(\bar{\mu}\,\Gamma\,\mu)(\bar{\chi}\,\Gamma'\,\chi),
	\end{align}
	where $\Gamma^{(')} \in \{1,\,\gamma^5,\,\gamma^\mu,\,\gamma^\mu\gamma^5,\,\sigma^{\mu\nu},\,\sigma^{\mu\nu}\gamma^5\}$.
	The full set of independent muonphilic operators, together with their
	leading direct detection (DD) scaling and DM annihilation properties, is
	shown in Table~\ref{tab:operators_muon}. The tensor operator $\bar{\psi}\sigma^{\mu\nu}\psi\bar{f}i\sigma_{\mu\nu}\gamma^5f$ can be written as $\bar{\psi}i\sigma^{\mu\nu}\gamma^5\psi\bar{f}\sigma_{\mu\nu}f\equiv\mathcal{O}_{pt}$ and $\bar{\psi}i\sigma^{\mu\nu}\gamma^5\psi\bar{f}i\sigma_{\mu\nu}\gamma^5f$ as $\bar{\psi}\sigma^{\mu\nu}\psi\bar{f}\sigma_{\mu\nu}f\equiv\mathcal{O}_{tt}$ by Fierz identities. So they are not treated separately. For the operators
	$\op_{ss}$, $\op_{pp}$, $\op_{sp}$, $\op_{ps}$, $\op_{tt}$ and $\op_{pt}$,
	these dimension-6 contact interactions in the WEFT (weak effective field theory) originate from
	dimension-7 SMEFT (SM effective field theory) operators\footnote{These dimension-7 operators can also induce semi-visible Higgs-boson decays at the LHC~\cite{Dawson:2025dmi} and mono-Higgs production at a muon collider~\cite{Belfkir:2023vpo}, whose dedicated investigation within the dimension-7 EFT framework is beyond the scope of the present study.} with a single Higgs doublet, so that their
	coefficients can be written as
	\begin{align}
	\mathcal{C}_i = \frac{v_h}{\Lambda}\,\mathcal{C}_i^{(7)}, \qquad
	i \in \{ss,pp,sp,ps,tt,pt\},
	\end{align}
	where, $v_h=246$ GeV is the Higgs vacuum expectation value. In our analysis we take this into account by switching on one operator
	at a time and setting the underlying $\mathcal{C}_i^{(7)}$ to unity. In
	this case the effective Lagrangian can be schematically written as
	\begin{align}
	\mathcal{L}
	= \mathcal{L}_{\rm SM}
	&+ \sum_{\substack{i\in\{vv,aa,\\av,va\}}}
	\frac{1}{\Lambda^2}\,\op_{i}
	+ \sum_{\substack{j\in\{ss,pp,sp,\\
			ps,tt,pt\}}}
	\frac{v_h}{\Lambda^3}\,\op_{j}\,.
	\end{align}

	\begin{table}[t]
		\caption{\small Dimension-6 operators coupling Dirac DM to muons, with their leading direct detection scaling~\cite{Fan:2010gt} and dominant annihilation partial wave. SI and SD denote spin-independent and spin-dependent scattering, $q$ is the momentum transfer, and $v$ is the relative velocity between DM and target.}
		\centering  
		\begin{tabular}{c c c c}
			\hline\hline
			Operator & Definition & Scattering  & Annihilation \\ 
			\hline\hline 
			$\mathcal{O}_{ss}$ & $\bar{\mu}\mu\,\bar{\chi}\chi$ & SI ~~ $(1)$ & $p$-wave\\  
			$\mathcal{O}_{pp}$ & $\bar{\mu}\gamma^5\mu\,\bar{\chi}\gamma^5\chi$ & SD ~~$(q^2)$ &$s$-wave\\
			$\mathcal{O}_{ps}$ & $\bar{\mu}i\gamma^5\mu\,\bar{\chi}\chi$& SD ~~$(q)$ &$p$-wave\\ 
			$\mathcal{O}_{sp}$ & $\bar{\mu}\mu\,\bar{\chi}i\gamma^5\chi$ & SI ~~$(q)$ &$s$-wave\\ 
			$\mathcal{O}_{vv}$ & $\bar{\mu}\gamma^{\mu}\mu\,\bar{\chi}\gamma_{\mu}\chi$ & SI ~~$(1)$ &$s$-wave\\
			$\mathcal{O}_{aa}$ & $\bar{\mu}\gamma^{\mu}\gamma^5\mu\,\bar{\chi}\gamma_{\mu}\gamma^5\chi$ & SD ~~$(1)$ &$s$-wave $\propto m_\mu^2/m_\chi^2$\\
			$\mathcal{O}_{av}$ & $\bar{\mu}\gamma^{\mu}\gamma^5\mu\,\bar{\chi}\gamma_{\mu}\chi$ & SD ~~$(v)$ &$s$-wave\\
			$\mathcal{O}_{va}$ & $\bar{\mu}\gamma^{\mu}\mu\,\bar{\chi}\gamma_{\mu}\gamma^5\chi$ & SD ~~$(v)$ &$p$-wave\\
			$\mathcal{O}_{tt}$ & $\bar{\mu}\sigma^{\mu\nu}\mu\,\bar{\chi}\sigma_{\mu\nu}\chi$ & SD ~~$(1)$ &$s$-wave\\
			$\mathcal{O}_{pt}$ & $\bar{\mu}i\sigma^{\mu\nu}\mu\,\bar{\chi}\sigma_{\mu\nu}\gamma^5\chi$ & SI ~~$(q)$ &$s$-wave\\
			\hline\hline
		\end{tabular}
		\label{tab:operators_muon}
	\end{table}
	
	Among these, $\mathcal{O}_{ss}$, $\mathcal{O}_{ps}$, $\mathcal{O}_{va}$ induce velocity-suppressed $p$-wave DM annihilation. Although $\mathcal{O}_{aa}$ induces $s$-wave annihilation, the rate is proportional to $m_\mu^2/m_\chi^2$ and thus suppressed for $m_\chi \gg m_\mu$. Other annihilation rates are $s$-wave and thus unsuppressed. The annihilation rates obtained from the operators are presented in the Appendix.

	\subsection{UV-completion: vector $L_\mu-L_\tau$ model}
	\label{subsec:vectormodel}
	
	As a first example of a UV-completion of muonphilic DM we consider the well-known gauged $U(1)_{L_\mu - L_\tau}$ model, which would generate the operator $\mathcal{O}_{vv}=\bar{\mu}\gamma^{\mu}\mu\,\bar{\chi}\gamma_{\mu}\chi$ in the $m_{\rm med}>\Lambda$ limit. For notational economy, let us denote $U(1)_{L_\mu - L_\tau}$ as $U(1)'$ from now on, with $Q' \equiv L_\mu - L_\tau$. In this setup, the SM is extended by a new gauge symmetry under which only muon- and tau-leptons (and their corresponding neutrinos) are charged, with opposite charges. A new gauge boson, $Z'$, mediates interactions between the DM and muonic current, naturally realising the muonphilic nature of the EFT. The relevant Lagrangian terms are
	\begin{align}
	\mathcal{L}_{L_\mu - L_\tau} = \ &\mathcal{L}_{\rm SM} - \frac{1}{4} Z^\prime_{\alpha\beta} Z'^{\alpha\beta}  + \frac{1}{2} m^2_{Z^\prime} Z^\prime_\alpha Z^{\prime \alpha} + \frac{\varepsilon}{2} Z^\prime_{\alpha\beta}  F^{\alpha\beta}  \nonumber\\
	& + g^{\prime} \big( \bar\mu \gamma^\alpha \mu - \bar\tau \gamma^\alpha \tau  
	+  \bar\nu_\mu \gamma^\alpha P_L\nu_\mu -  \bar\nu_\tau \gamma^\alpha P_L \nu_\tau\big) Z'_\alpha \nonumber\\
	&+g^{\prime} Q^{\prime}(\chi)\,[\bar{\chi}\gamma^\alpha\chi] \,Z_\alpha^\prime\ , 
	\end{align}
	where $Z'_{\alpha}$ is the $U(1)'$ gauge boson, with field strength $Z'_{\alpha\beta} = \partial_\alpha Z'_\beta - \partial_\beta Z'_\alpha$, $m_{Z'}$ is its mass (arising from a dark Higgs or Stückelberg mechanism), $g^{\prime}$ is the gauge coupling of the muon–$Z'$ interaction (with opposite sign coupling to tau leptons), $g_\chi=g^{\prime} Q^{\prime}(\chi)$ is the DM–$Z'$ gauge coupling, implementing the $Q' = L_\mu - L_\tau$ charge assignment, and $m_\chi$ is the mass of the DM. This setup introduces several new parameters $m_{Z'}$, $g_\chi$ and $g^{\prime}$ as well as $m_\chi$. The kinetic mixing parameter $\varepsilon$ receives a momentum dependent loop-contribution apart from the tree-level value $\varepsilon_0$, with asymptotic limits~\cite{Hapitas:2021ilr},
	\begin{align}
	\varepsilon(Q^2&\gg m_{\tau}^2)= \varepsilon_0\nonumber\\
	\varepsilon(Q^2&\ll m_{\mu}^2)= \varepsilon_0 -\frac{e\,g^{\prime}}{12\pi^2} \log \left(\frac{m_{\tau}^2}{m_{\mu}^2}\right).\nonumber
	\end{align}
	For our muon collider study the $Q^2\gg m_{\tau}^2$ limit holds, and we choose $\varepsilon_0=0$ to suppress kinetic mixing effects.
	
	In the kinematic regimes relevant for relic-density and ADM considerations the dominant DM annihilation channels in the $U(1)'$ model are
	$\chi\bar\chi\to \ell^+\ell^-$ (visible), $\chi\bar\chi\to \nu_\ell\bar\nu_\ell$ (invisible) (with $\ell = \mu,\tau$) and, when kinematically open, $\chi\bar\chi\to Z'Z'$.  The leading order (non-relativistic) expressions for the annihilation rates $\langle\sigma v\rangle$ for this model are presented in the Appendix.
	
	\subsection{UV-completion: axial ${L_\mu-L_\tau}$ model}
	\label{subsec:axialvectormodel}
	
	We now present a second, minimal UV completion that reproduces an $\mathcal{O}_{va}=\bar{\mu}\gamma^{\mu}\mu\,\bar{\chi}\gamma_{\mu}\gamma^5\chi$ effective operator at low energies while naturally accommodating asymmetric relic dynamics.\footnote{This model is one possible UV completion of the scenario analysed in Ref.\cite{Wang:2025cth}.} Following the previous $L_\mu-L_\tau$ model, we extend the SM field content by the $U(1)'$ gauge boson $Z'$, but now with two SM-singlet Dirac fermions $\chi,\psi$, and a complex singlet scalar $S$, and assign chiral $U(1)'$ charges to the dark-sector fermions such as to make the model anomaly-free:
	\begin{align}
	Q'(\mu_L)=Q'(\mu_R)=+1,\quad
	Q'(\tau_L)=Q'(\tau_R)=-1,
	\end{align}
	and
	\begin{align}
	Q'(\chi_L)=+q,\quad Q'(\chi_R)=-q,\nonumber\\
	Q'(\psi_L)=-q,\quad Q'(\psi_R)=+q,\nonumber\\
	Q'(S)=Q'(\chi_L)-Q'(\chi_R)=2q.
	\end{align}
	The renormalisable Lagrangian is
	\begin{align}
	\mathcal L \;=\;
	&\ \mathcal L_{\rm SM}
	-\frac14 Z'_{\alpha\beta}Z'^{\alpha\beta}
	+\frac12 m_{Z'}^2 Z'_\alpha Z'^{\alpha}
	+\frac{\varepsilon_0}{2} Z'_{\alpha\beta}F^{\alpha\beta}
	\nonumber\\
	&\ +\, g'\,Z'_\alpha\!\Big(
	\bar\mu\gamma^\alpha\mu-\bar\tau\gamma^\alpha\tau
	+\bar\nu_\mu\gamma^\alpha P_L\nu_\mu-\bar\nu_\tau\gamma^\alpha P_L\nu_\tau
	\Big)
	\nonumber\\
	&\ +\, g'\,Z'_\alpha\left(
	q\,\bar\chi\gamma^\alpha\gamma^5\chi \;-\; q\,\bar\psi\gamma^\alpha\gamma^5\psi
	\right)
	\nonumber\\
	&\ -\Big(y_\chi S\,\bar\chi_L\chi_R + y_\psi S^\dagger \bar\psi_L\psi_R + \text{h.c.}\Big)
	\;-\;\Big(\bar\chi_L\,m_{12}\,\psi_R + \bar\psi_L\,m'_{12}\,\chi_R + \text{h.c.}\Big)
	\nonumber\\
	&\ +\,|\partial_\alpha S + i g' Q'(S) Z'_\alpha S|^2 - V(H,S).
	\label{eq:lag-mixing}
	\end{align}
	Note that the parameters $y_{\chi}$, $y_{\psi}$, $m_{12}$ and $m'_{12}$ can be made real by rephasing the four fermion fields $\chi_{L,R}$ and $\psi_{L,R}$. To simplify the analysis, we also choose the benchmark parameter point $m'_{12} = m_{12}$ in what follows.
	
	After $S$ acquires a nonzero vacuum expectation value (VEV), $\langle S \rangle = v_S/\sqrt{2}$, the $Z'$ obtains the mass
	\begin{equation}
	m_{Z'} = 2\, q\, g'\, v_S.
	\end{equation}
	Also, the $\chi$ and $\psi$ fermion mass matrix $\mathbf{M}$ in
	\begin{equation}
	\left( \begin{array}{cc} \bar{\chi}_L & \bar{\psi}_L \end{array} \right) \mathbf{M} \left( \begin{array}{c} \chi_R \\ \psi_R \end{array} \right) + h.c.
	\end{equation}
	is generated and given by
	\begin{equation}
	\mathbf{M} =
	\begin{pmatrix}
	m_\chi & m_{12} \\
	m_{12} & m_\psi
	\end{pmatrix}
	\end{equation}
	where
	\begin{align}
	m_\chi = \frac{y_\chi v_S}{\sqrt{2}}\quad \text{and}\quad
	m_\psi = \frac{y_\psi v_S}{\sqrt{2}}. 
	\end{align}
	The mixing angle and eigenvalues are
	\begin{align}
	\tan(2\theta) = \frac{2 m_{12}}{m_\chi - m_\psi},
	\qquad
	m_{1,2} = \frac{1}{2} \Big[ (m_\chi + m_\psi) \mp \sqrt{(m_\chi - m_\psi)^2 + 4 m_{12}^2} \Big].
	\end{align}
	Since $\mathbf{M}$ is real symmetric, the same orthogonal rotation $U(\theta)$ transforms both chiralities into the mass basis:
	\begin{align}
	\begin{pmatrix} \chi_L \\ \psi_L \end{pmatrix}
	= U(\theta) \begin{pmatrix} X_{1L} \\ X_{2L} \end{pmatrix},
	\qquad
	\begin{pmatrix} \chi_R \\ \psi_R \end{pmatrix}
	= U(\theta) \begin{pmatrix} X_{1R} \\ X_{2R} \end{pmatrix},
	\qquad
	U(\theta) =
	\begin{pmatrix}
	\cos\theta & \sin\theta \\
	-\sin\theta & \cos\theta
	\end{pmatrix},
	\end{align}
	with $X_{1,2}$ being the mass eigenstate fields.
	
	The $Q'$ charge matrices for the LH and RH fermions are
	\begin{align}
	Q'_L =
	\begin{pmatrix}
	q & 0 \\
	0 & -q
	\end{pmatrix},
	\qquad
	Q'_R =
	\begin{pmatrix}
	-q & 0 \\
	0 & q
	\end{pmatrix},
	\end{align}
	respectively. Thus the vector combination $Q'_V \equiv (Q'_L + Q'_R)/2$ vanishes, while the axial combination is $Q'_A \equiv (Q'_L - Q'_R)/2 = \mathrm{diag}(q, -q)$, with
	\begin{align}
	U^\dagger Q'_A U
	= q
	\begin{pmatrix}
	\cos(2\theta) & \sin(2\theta) \\
	\sin(2\theta) & -\cos(2\theta)
	\end{pmatrix}.
	\end{align}
	This leads to the $Z'$ interaction being purely axial:
	\begin{align}
	\mathcal{L}_{Z'}^{\text{mass}}
	\supset g' Z'_\alpha \Big[
	q \cos(2\theta) \big( \bar{X}_1 \gamma^\alpha \gamma^5 X_1 - \bar{X}_2 \gamma^\alpha \gamma^5 X_2 \big)
	+ q \sin(2\theta) \big( \bar{X}_1 \gamma^\alpha \gamma^5 X_2 + \bar{X}_2 \gamma^\alpha \gamma^5 X_1 \big)
	\Big].
	\label{eq:Zprime-axial}
	\end{align}
	
	We take $X_1$ as the DM with $m_{X_1}<m_{X_2}$, and define
	\begin{align}
	\label{eq:gax}
	g_{\text{ax}} \equiv g' q \cos(2\theta).
	\end{align}
	The annihilation rates for this model are similar to the vector $U(1)_{\lmultau}$ case, except for an extra velocity dependent factor in the $X_1 X_1\to \ell^+ \ell^-$ channel arising from the axial-vector coupling. The relevant expressions are given in the Appendix.
	
	\section{Relic density of ADM}
	\label{sec:relicdensity}
	
	The ADM paradigm has both DM ($\chi$) and anti-DM ($\bar{\chi}$) existing in the early Universe, analogous to baryons and antibaryons. A primordial asymmetry between these sectors is generated through some mechanism, which freezes in once thermal equilibrium is lost. The symmetric component then needs to annihilate away efficiently, leaving only the asymmetric relic. In this work, we do not specify the asymmetry generation dynamics, including how the chemical equilibration of the baryon and dark asymmetries occurs. We expect there to be several ways of constructing these dynamics. We only need to assume that those dynamical choices do not affect our analysis.\footnote{We also assume that there is no late entropy injection (or entropy production) and no dark-sector temperature difference that could modify the relic density.}

	The total DM yield can be expressed as
	\begin{align}
	Y_{\rm tot} \;=\; Y_{\chi} + Y_{\overline{\chi}}
	\;=\; (Y_{\chi} - Y_{\overline{\chi}}) + 2 Y_{\overline{\chi}}
	\;=\; Y_{\rm asy} + Y_{\rm sym},
	\end{align}
	where $Y \equiv n/s$ is the number density $n$ normalised by the entropy density $s$ of the universe. Here $Y_{\rm asy}$ denotes the asymmetric yield and $Y_{\rm sym}$ the symmetric one~\cite{Graesser:2011wi,Iminniyaz:2011yp}.
	
	Following~\cite{Iminniyaz:2011yp}, the symmetric contribution is
	\begin{align}
	Y_{\rm sym} \;=\; 2Y_{\overline{\chi}} \;=\; 
	\frac{2 Y_{\rm asy}}{
		\exp\left[ Y_{\rm asy} \lambda \left( \frac{a}{x_F} + \frac{3b}{x_F^2} \right) \right] - 1},
	\label{eq:ydmbar}
	\end{align}
	where $a$ and $b$ are the coefficients of the $s$-wave and $p$-wave terms in the partial-wave expansion $\langle\sigma v\rangle = a + b v^2 + \dots$ of the annihilation cross-section, and
	\begin{align}
	\lambda \;=\; \frac{4\pi}{\sqrt{90}} \, m_{\chi} M_{\rm pl} \sqrt{g_*},
	\end{align}
	with $M_{\rm pl}=2.4\times 10^{18}\,\mathrm{GeV}$ the reduced Planck mass and $g_* \simeq 100$ the relativistic degrees of freedom at freeze-out. The modified freeze-out parameter is~\cite{Iminniyaz:2011yp}
	\begin{align}
	x_F \;=\; x_{F_0} \left[ 1 + 0.285\,\frac{a \lambda Y_{\rm asy}}{x_{F_0}^3} 
	+ 1.35\,\frac{b \lambda Y_{\rm asy}}{x_{F_0}^4} \right],
	\label{eq:xf}
	\end{align}
	where $x_{F_0} = m_\chi / T_F$ is the usual WIMP freeze-out value. The shift $x_F - x_{F_0}$ encodes the ADM correction to the $\bar{\chi}$ abundance, absent in the symmetric case. As noted earlier, the annihilation cross sections $\langle\sigma v\rangle$ for the EFT operators in Table~\ref{tab:operators_muon} and the two UV models are given in the Appendix.
	
	We require that the symmetric part in Eq.~\eqref{eq:ydmbar} be subdominant, e.g.
	\begin{align}
	Y_{\rm sym} \;\leq\; \frac{1}{100} \times 
	\frac{\Omega_{\rm DM} h^2}{2.76\times 10^8}
	\left( \frac{\mathrm{GeV}}{m_\chi} \right),
	\label{eq:constraint}
	\end{align}
	corresponding to $\lesssim 1\%$ of the observed dark matter density~\cite{March-Russell:2012elz}. We term this the \textit{ADM condition}. Efficient annihilation into SM fermions via the operators of Table~\ref{tab:operators_muon} can satisfy this requirement, but also faces limits from laboratory and astrophysical probes. For a given $m_\chi$, this results in an upper bound $\Lambda < \Lambda_1$ from the ADM condition, and a lower bound $\Lambda > \Lambda_2$ from null searches. The allowed region requires $\Lambda_1 > \Lambda_2$; if $\Lambda_1 \leq \Lambda_2$ in some mass range, the ADM scenario is excluded there. Operators with $p$-wave annihilation, such as $\mathcal{O}_{ss}$, $\mathcal{O}_{ps}$, and $\mathcal{O}_{va}$, demand smaller $\Lambda$ to satisfy Eq.~\eqref{eq:constraint}, leading to stronger exclusions. The same logic applies to the UV models, with the role of $\Lambda$ played by appropriate combinations of mediator mass and couplings, such as $m_{Z'}/\sqrt{g^{\prime}g_\chi}$.
	
	\section{Signal at a muon collider}
	
	\label{sec:muoncollider}
	
\begin{figure}
		\centering
		\begin{subfigure}[b]{0.3\textwidth}
			\centering
			\begin{tikzpicture}
			\begin{feynman}
			\vertex (a) at (-2.0, 1.6) {$\mu^-$};
			\vertex (b) at (-1.8, -1.4) {$\mu^+$};
			\vertex (c1) at ( 1.8, 1.4) {$\nu$};
			\vertex (d) at ( 1.8, -1.4) {$\bar{\nu}$};
			\vertex (q) at (-0.1,1.2) {$\gamma$};
			\vertex[dot, scale=0.7] (p) at (-1.2,0.6) {};
			\vertex[dot, scale=0.7] (e) at (-0.75, 0){};
			\vertex[dot, scale=0.7] (f) at ( 0.75, 0) {\contour{white}{}};
			\diagram* {
				(a) -- [fermion] (p) -- [fermion] (e) -- [fermion] (b),
				(p) -- [photon] (q),
				(d) -- [fermion] (f) -- [fermion] (c1),
				(e) -- [photon, edge label=$Z/\gamma$] (f)
			};
			\end{feynman}
			\end{tikzpicture}
			\caption*{(SM)}
		\end{subfigure}
		\begin{subfigure}[b]{0.3\textwidth}
			\centering
			\begin{tikzpicture}
			\begin{feynman}
			\vertex (a) at (-2.0, 1.6) {$\mu^-$};
			\vertex (b) at (-1.8, -1.4) {$\mu^+$};
			\vertex (c1) at ( 1.8, 1.4) {$\chi$};
			\vertex (d) at ( 1.8, -1.4) {$\bar{\chi}$};
			\vertex (q) at (-0.1,1.2) {$\gamma$};
			\vertex[dot, scale=0.7] (p) at (-1.2,0.6) {};
			\vertex[blob, scale=0.5] (e) at (-0.75, 0){};
			\vertex[blob, scale=0.5] (f) at ( 0.75, 0) {\contour{white}{}};
			\diagram* {
				(a) -- [fermion] (p) -- [fermion] (e) -- [fermion] (b),
				(p) -- [photon] (q),
				(d) -- [fermion] (f) -- [fermion] (c1),
				(e) -- [photon, edge label=$Z^\prime$] (f)
			};
			\end{feynman}
			\end{tikzpicture}
			\caption*{(UV)}
		\end{subfigure}
		\begin{subfigure}[b]{0.3\textwidth}
			\centering
			\begin{tikzpicture}
			\begin{feynman}
			\vertex (a) at (-1.5, 1.6) {$\mu^-$};
			\vertex (b) at (-1.3, -1.4) {$\mu^+$};
			\vertex (c) at ( 1.3, 1.4) {$\chi$};
			\vertex (d) at ( 1.3, -1.4) {$\bar{\chi}$};
			\vertex (q) at (0.5,1.2) {$\gamma$};
			\vertex[dot, scale=0.7] (p) at (-0.6,0.6) {};
			\vertex[blob, scale=0.5] (e) at (0, 0) {\contour{white}{}};
			\diagram* {
				(a) -- [fermion] (p) -- [fermion] (e) -- [fermion] (b),
				(p) -- [photon] (q),
				(d) -- [fermion] (e) -- [fermion] (c)
			};
			\end{feynman}
			\end{tikzpicture}
			\caption*{(EFT)}
		\end{subfigure}
		\caption{\small A few representative Feynman diagrams of the mono-photon production process in the SM (left), UV-completions with $Z^\prime$ mediator (centre), and EFT (right) at a muon collider. The blobs represents new physics interactions.}
		\label{fig:monojet_feynman}
	\end{figure}
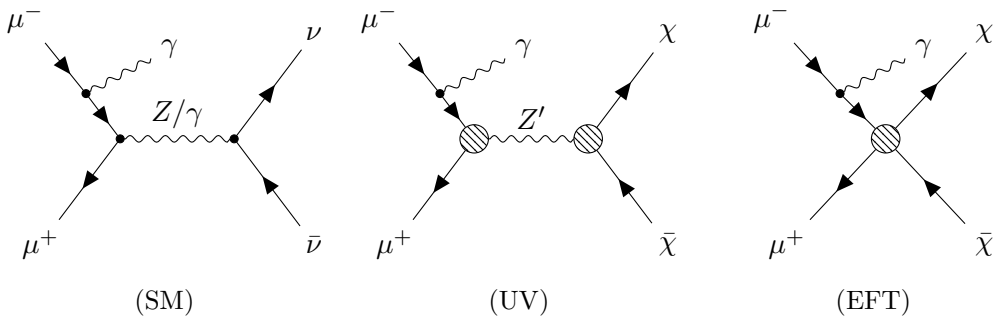

	We now study the sensitivity of a high energy muon collider to muonphilic ADM through the mono–photon channel,
	\begin{align}
	\mu^+ \mu^- \to \chi \bar{\chi} \gamma.
	\end{align}
	The photon originates from initial state radiation (ISR) and the DM pair escapes undetected, resulting in missing transverse momentum, with the photon providing the experimental handle to trigger on and reconstruct the event. In Fig.~\ref{fig:monojet_feynman} we show some of the Feynman diagrams for this process both for the EFT framework and the UV-models.
	
	The models are written in \texttt{FeynRules}~\cite{Alloul:2013bka} to implement the relevant interactions and exported to a \texttt{UFO} model. This \texttt{UFO}, in turn, is imported to \texttt{MadGraph5\_aMC@NLO}~\cite{Alwall:2014hca} for parton–level generation, followed by \texttt{Pythia8}~\cite{Sjostrand:2006za,Sjostrand:2007gs} for an approximate treatment of the initial state radiation, and \texttt{Delphes}~\cite{deFavereau:2013fsa} with the muon collider detector card to account for detector resolution and acceptance. We consider two benchmark collider setups: $\sqrt{s}=3~\mathrm{TeV}$ and $10~\mathrm{TeV}$, each with an integrated luminosity of $1~\mathrm{ab}^{-1}$ and unpolarised beams \cite{InternationalMuonCollider:2024jyv}.
	
\begin{figure}
		\centering
		\includegraphics[width=0.49\linewidth]{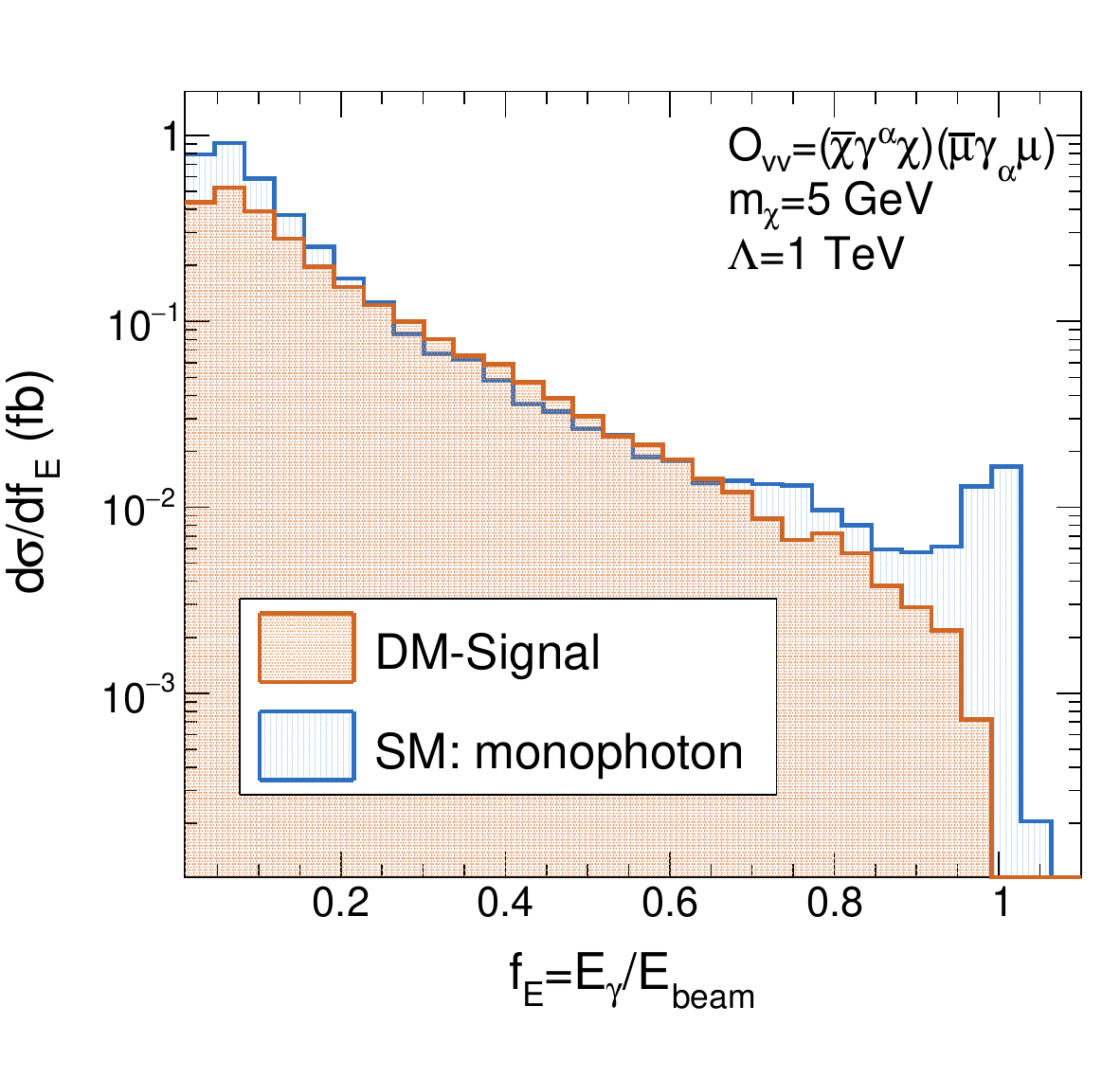}
		\includegraphics[width=0.49\linewidth]{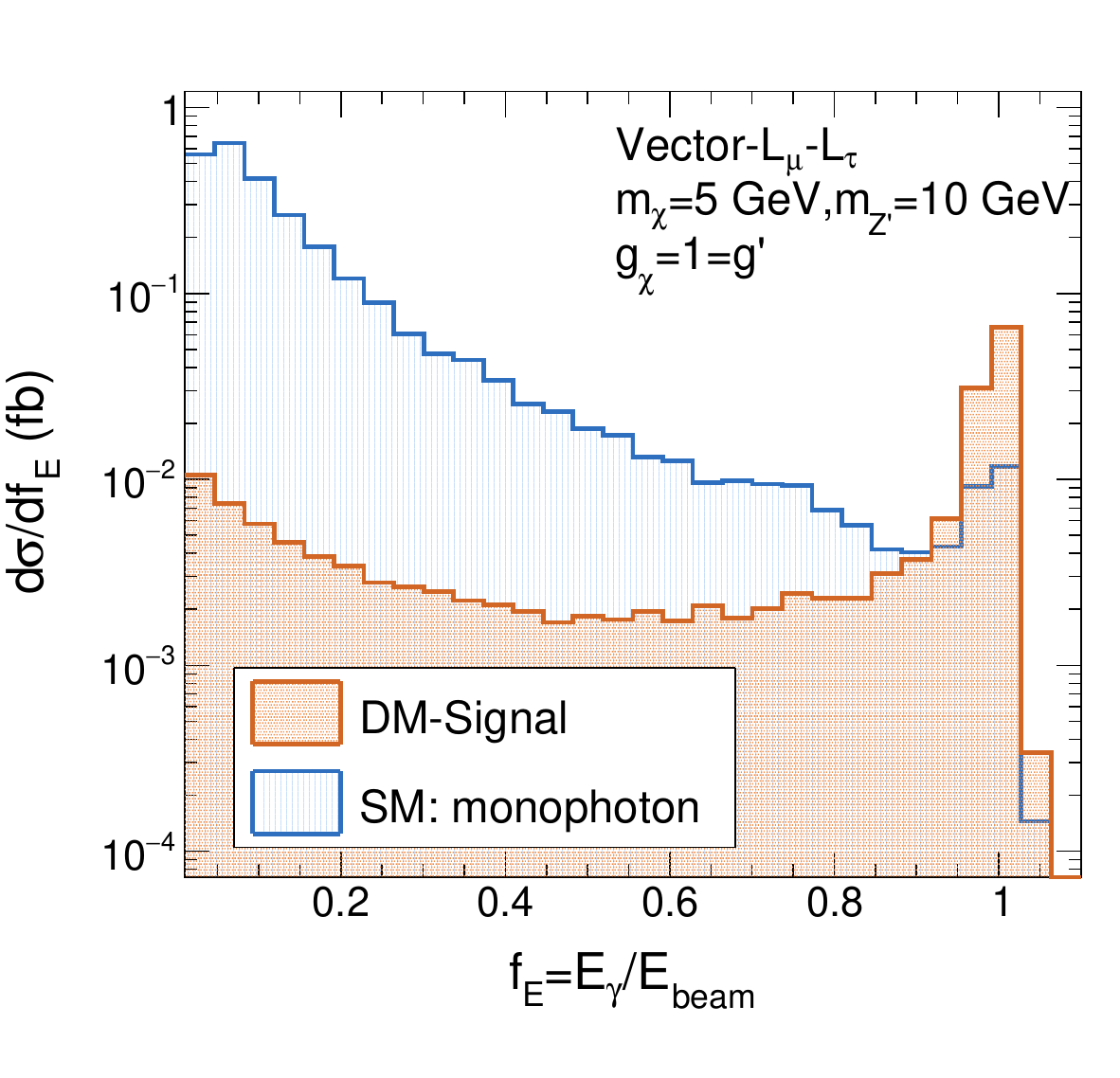}
		\caption{Distribution of the photon energy fraction in mono-photon events. The left panel corresponds to the EFT case for $\op_{vv}$, while the right panel displays the vector-like UV model. For the EFT case we choose the benchmark values $\rm \Lambda=1~TeV$ and $m_\chi=5$ GeV, while for the UV case we use $g_\chi=1$, $g^{\prime}=1$, $m_\chi=5$ GeV and $m_{Z^\prime}=10$ GeV.}
		\label{fig:photon_Efrac}
	\end{figure}

	The dominant irreducible background arises from the SM process
	\begin{align}
	\mu^+ \mu^- \to \nu \bar{\nu} \gamma,
	\end{align}
	with $\nu = \nu_\mu,\nu_\tau$ produced via $t$–channel $W$ or $s$–channel $Z$ exchange in the SM, with the latter depicted in the leftmost diagram of Fig.~\ref{fig:monojet_feynman}. This background has the same visible final state as the signal, but a slightly different photon energy spectrum: ISR photons from SM neutrino production have a more forward peak than for DM production, coming from the on-shell Z-mediation, as we see in Fig.~\ref{fig:photon_Efrac}. As the center-of-mass energy $\sqrt{s}$ moves away from the Z pole, this peak becomes increasingly suppressed, and eventually the SM process becomes an almost irreducible background to the mono-photon signal from DM events. Thus, a muon collider running close to the Z pole would be more efficient at suppressing this dominant SM background. Nevertheless, as we will show later, the present muon collider configuration remains sensitive to these scenarios.
	
	To suppress the background and enhance the signal sensitivity, we select events containing exactly one isolated photon with transverse momentum $p_T^\gamma$ above a given threshold and pseudorapidity $|\eta^\gamma|$ within the detector acceptance. Variables such as missing transverse energy and photon energy are highly correlated with the photon energy fraction ($f_E$), so the selection $f_E < 0.8$, turns out to be optimal for the EFT scenario. For the $L_{\mu}-L_\tau$ model, however, the $Z^\prime$ can be produced on shell and propagates in the opposite direction to the photon to conserve momentum. The photon recoil creates a peak at $f_E\sim 1$. The size of this on-shell peak depends on the model parameters. Due to this feature, the optimal selections is opposite to the EFT: $f_{E}>0.8$.
	
	The energy dependence of the monophoton signal is qualitatively
	different in the EFT and UV regimes. In the EFT case,
	$m_{Z'}^2\gg \hat s$, the propagator reduces to a contact interaction
	so that the inclusive monophoton cross section grows approximately like
	$\sigma_{\mu^+\mu^-\to\chi\bar\chi\gamma}\propto s/\Lambda^4$ (up to
	logarithmic ISR enhancements). In contrast, once the collider energy is
	taken well above the mediator mass, $\hat s\gg m_{Z'}^2$ over most of
	the phase space and the off-shell propagator implies
	$\sigma_{\mu^+\mu^-\to\chi\bar\chi\gamma}\propto 1/s$. Since the mediator
	masses of interest are already accessible at a $\sqrt{s}=3~\text{TeV}$
	muon collider, a $\sqrt{s}=10~\text{TeV}$ machine would mainly probe the
	high-energy off-shell tail of the $Z'$ in the UV model, where the signal
	yield is reduced. For this reason we present UV benchmarks only for the
	$3~\text{TeV}$ option, and use the $10~\text{TeV}$ stage to explore the
	EFT regime only. The $10~\text{TeV}$ muon collider would, however, be important to probe larger $Z^\prime$ masses, with $m_{DM}\sim 5$ TeV.
	
	The statistical sensitivity is evaluated using the total photon energy spectrum, through the measure $\rm S/\sqrt{S+B}$, where $S$ ($B$) is the number of signal (background) events. The $3\sigma$ reach in the $(m_\chi,\Lambda)$ plane for the EFT and in the $(m_{Z'}, g^{\prime})$ plane (with $m_\chi=5$ GeV and different values of $g_{\chi}$) for the $L_\mu-L_\tau$ models are extracted. These muon collider projections are then combined with the astrophysical capture, other collider limits, and ADM requirements discussed in earlier sections to determine the muon collider parameter reach. The results are presented in Figs.~\ref{fig:EFT_exclusion1}-\ref{fig:EFT_exclusion5}  for the EFT cases and Figs.~\ref{fig:UV_exclusion} and \ref{fig:UV_exclusion_axial} for the $L_\mu-L_\tau$ models, together with existing bounds that we now discuss.

	\section{Other bounds}
	\label{sec:other_bounds}
	
	Before presenting the muon-collider sensitivity reach in Sec.~6 (Figs.~\ref{fig:EFT_exclusion1}--\ref{fig:UV_exclusion_axial}), in this section we first compute the direct-detection bounds on the EFT operators and UV models, and summarise the other constraints that feed into those figures. Note that these exclusions do not rely on the ADM assumption: they apply to any DM candidate with the EFT interactions in Table~\ref{tab:operators_muon} or the UV completions considered here.
	
	\subsection{Direct detection}
	\begin{table}[t]
		\caption{\small Loop-induced WIMP--nucleus cross sections for muonphilic dimension-6 operators following Ref.~\cite{Kopp:2009et}. Entries marked ``0'' vanish at this order because the muon current is axial or pseudoscalar.}
		\centering
		\begin{tabular}{c l}
			\hline\hline
			Operator & $\sigma(\chi N \to \chi N)$ \\
			\hline\hline
			$\mathcal{O}_{ss}$  &
			$ 2\left( \frac{\alpha_{\rm em} Z}{\pi} \right)^2 \sigma^{(2)}_N
			\left( \frac{\pi^2}{12} \right)^2 \left( \frac{\mu_N v}{m_\mu} \right)^2 \quad$ (2-loop) \\
			$\mathcal{O}_{pp}$ & $0$ \\
			$\mathcal{O}_{ps}$ & 0\\
			$\mathcal{O}_{sp}$  &
			$ \frac{4}{3}\left( \frac{\alpha_{\rm em} Z}{\pi} \right)^2 \sigma^{(2)}_N
			\left( \frac{\pi^2}{12} \right)^2 \left( \frac{\mu_N^2 v^2}{m_\mu m_\chi} \right)^2\quad$ (2-loop) \\
			$\mathcal{O}_{vv}$ &
			$ \frac{1}{9} \sigma^{(1)}_N L_\mu^2  \quad$ (1-loop) \\
			$\mathcal{O}_{aa}$ & $0$ \\
			$\mathcal{O}_{av}$  & $0$ \\
			$\mathcal{O}_{va}$  &
			$ \sigma^{(1)}_N L_\mu^2 \frac{v^2}{9}
			\left( 1 + \frac{1}{2} \frac{\mu_N^2}{m_N^2} \right) \quad$ (1-loop) \\
			$\mathcal{O}_{tt}$  &
			$ \sigma^{(2)}_N L_\mu^2 \frac{m_\mu^2}{\mu_N^2}
			\ln \frac{E^{\max}_d}{E^{\min}_d} \quad$ (1-loop) \\
			$\mathcal{O}_{pt}$ &
			$ \sigma^{(2)}_N L_\mu^2 \frac{m_\mu^2}{\mu_N^2}
			\frac{1}{v^2} \ln \frac{E^{\max}_d}{E^{\min}_d} \quad$ (1-loop) \\
			\hline\hline
		\end{tabular}
		\label{tab:DD_muon_loops}
	\end{table}
	
	\paragraph{EFT operators:} 
	Tree level DM--nucleon interactions are absent in our muonphilic EFT setup: the operators in Table~\ref{tab:operators_muon} couple DM to leptons only, and the SM quarks are neutral under these contact terms. Nevertheless, loop-induced couplings to nuclei do arise once a photon is attached to the charged-lepton line, effectively matching the leptonic current onto the nuclear electromagnetic current. In the heavy-mediator (contact) regime relevant for nuclear recoils, this generates a coherent amplitude proportional to atomic number $Z$, while preserving the operator-specific velocity and mass suppressions. Defining
	\begin{align}
	\sigma_N^{(1)} = \frac{\mu_N^2}{\pi} \left( \frac{\alpha_{\rm em} Z}{\pi \Lambda^2} \right)^2,
	\qquad \sigma_N^{(2)} = \frac{\mu_N^2}{\pi} \left( \frac{\alpha_{\rm em} Z\, v_h}{\pi \Lambda^3} \right)^2,\qquad
	L_\mu^2 = \ln^2 \left( \frac{m_\mu^2}{\Lambda^2} \right),
	\end{align}
	with $\mu_N$ being the DM-nucleus reduced mass, and $E_d^{\min,\max}$ denoting the recoil-energy analysis window $E_d\sim [1-50]~\text{KeV}$ \cite{LZ:2024zvo}, the expressions of the loop-induced scattering cross-sections $\sigma(\chi N \to \chi N)$ are collected in Table~\ref{tab:DD_muon_loops}~\cite{Kopp:2009et}. At this order, axial and pseudoscalar \emph{muon} currents do not generate a coherent $\chi N$ amplitude (their loop matching to the nuclear EM current vanishes), while vector and tensor structures do, with the velocity and mass dependences indicated in the table. In our numerical limits we adopt the experimental recoil ranges and exposures of  PandaX-4T~\cite{PandaX:2022xqx, PandaX:2023xgl} and LZ~\cite{LZ:2024zvo} for the spin-independent cases, while Pico-60~\cite{PICO:2019vsc} and LZ~\cite{LZ:2024zvo} are used for the spin-dependent cross sections. We combine the constraints from different experiments consistently for the mass ranges indicated in the main text.

	\paragraph{Vector $U(1)_{L_\mu - L_\tau}$ DM current:}
	In the heavy-mediator regime relevant for nuclear recoils ($q^2 \ll m_{Z'}^2$), the $t$-channel $Z'$ exchange reduces to a contact interaction and the spin-independent cross section reads~\cite{Altmannshofer:2016jzy}
	\begin{align}
	\sigma_{\chi N} = \frac{1}{A^2} \frac{\mu_{\chi N}^2}{9 \pi}
	\left( \frac{\alpha_{\rm em} g^{\prime} g_{\chi}}{\pi m_{Z'}^2}
	\ln \frac{m_\mu^2}{m_\tau^2} \right)^2 Z^2,
	\label{eq:dd_lmultau}
	\end{align}
	with $\mu_{\chi N}$ the DM--nucleus reduced mass and $g^{\prime}$, $g_{\chi}$ the $Z'$ couplings to the muon current and to DM, respectively. Since $q \simeq \sqrt{2 m_N E_R} \lesssim \mathcal{O}(10~\text{MeV})$ for $E_R \sim \text{keV}$, the contact limit is valid for $m_{Z'} \gtrsim \mathcal{O}(\text{GeV})$. We use the spin-independent DD limits in this case.
	
	\paragraph{Axial $U(1)_{L_\mu - L_\tau}$ DM current:}
	Muon and tau loops induce an effective momentum-independent $Z'$--photon mixing in the recoil regime, yielding
	\begin{align}
	\mathcal{L}_{\rm eff} = C (\bar{X}_1 \gamma^\mu \gamma^5 X_1) J^{\rm EM}_\mu,
	\qquad
	C = \frac{\alpha_{\rm em}}{\pi} \frac{g' g_{\rm ax}}{m_{Z'}^2}
	\ln \frac{m_\mu^2}{m_\tau^2},
	\end{align}
	where $g_{\rm ax} = g' q \cos(2\theta)$. Matching to the non-relativistic limit~\cite{Fan:2010gt},
	\begin{align} 
	\langle N | J^{0}_{\rm EM} | N \rangle \simeq Z, \qquad \bar{u}(p') \gamma^0 \gamma^5 u(p) \to 2 \boldsymbol{S}_\chi \cdot \boldsymbol{v}^\perp, \qquad \big\langle (\boldsymbol{S}_\chi \cdot \boldsymbol{v}^\perp)^2 \big\rangle = \frac{v^2}{12}, 
	\end{align}
	where $\boldsymbol{S}_\chi$ is the DM spin operator, and $\boldsymbol{v}^\perp$ is the
	(transverse) DM-nucleon relative velocity, respectively, one obtains the elastic cross section
	\begin{align}
	\sigma_{X_1 N}^{(\rm elastic)} =
	\frac{1}{A^2} \frac{\mu_{X_1 N}^2}{\pi} \frac{v^2}{12}
	\left( \frac{\alpha_{\rm em} g' g_{\rm ax}}{\pi m_{Z'}^2}
	\ln \frac{m_\mu^2}{m_\tau^2} \right)^2 Z^2,
	\label{eq:sigma-elastic-axial}
	\end{align}
	which is the axial analogue of Eq.~\eqref{eq:dd_lmultau}, differing by the replacement $g_{\chi} \to g_{\rm ax}$ and an additional $v^2$ suppression from the axial--vector non-relativistic matching. The spin-dependent DD constraints are used in this case.

	\subsection{DM capture in neutron stars}
	\label{sec:NS}
	Dark matter capture in neutron stars (NS) have interesting prospects to probe leptophilic DM in the MeV–TeV mass range~\cite{1985ApJ296679P,1987ApJ321560G,1987ApJ321571G,Bertone:2007ae,Bramante:2017xlb,Dasgupta:2019juq,Dasgupta:2020dik}. Owing to their extreme densities and large escape velocities, NS can efficiently capture DM through scattering on leptons present in their interiors. For muonphilic scenarios, this is particularly relevant because, in addition to highly degenerate electrons, the cores of NS can host a significant muon population, enabling efficient capture via DM–muon scattering~\cite{Kouvaris:2007ay,Kouvaris:2010vv,Baryakhtar:2017dbj,Bell:2018pkk,Chen:2018ohx,Acevedo:2019agu,Raj:2017wrv,Bramante:2017xlb,deLavallaz:2010wp,Bell:2019pyc,Bell:2020lmm,Bell:2021fye}. Since old, isolated NS are expected to cool well below 1000 K, the absence of anomalously warm NS in observations will translate into stringent upper limits on the DM–muon scattering cross section. We adopt the results of NS capture for the EFT interactions from~\cite{Bell:2020lmm} considering the BSk24-2 NS benchmark model. For the specific gauged vector-$U(1)_{\lmultau}$ model, projected neutron-star sensitivities have been studied in detail for benchmark mediator masses and couplings, mainly in the light-mediator regime~\cite{Bell:2025acg,Garani:2019fpa}. A dedicated re-evaluation of these sensitivities for our UV benchmarks is beyond the scope of this work.

	\subsection{Collider bounds}
	For the $\lmultau$ model, high-energy colliders provide a powerful and complementary way to probe the interaction strengths. At the LHC, a $Z^{\prime}$ mediator can be radiated from muons produced in Drell–Yan or Z-boson decays, and subsequently decay back into a dimuon pair. This leads to clean four-$\mu$ final states with small SM backgrounds and excellent invariant-mass resolution, making them an ideal laboratory for testing muonphilic scenarios.
	
	In our analysis, we make use of the CMS constraints on a light $Z^{\prime}$ associated with a gauged $U(1)_{\lmultau}$ symmetry, performed in the $pp\to Z\to 4\mu$ channel at $\sqrt{s}=13$ TeV~\cite{CMS:2018yxg}. The search targets the topology $pp\to Z^{\prime}\mu^+\mu^-, \,Z^{\prime}\to \mu^+\mu^-$, and looks for a narrow resonance in the dimuon invariant mass spectrum on top of the SM $Z\to 4\mu$ background. Assuming a pure $\lmultau$ model, which means no coupling of $Z^{\prime}$ to DM ($\chi$), Ref.~\cite{CMS:2018yxg} has put direct constraints on the regions of ($m_{Z^{\prime}},g^{\prime}$) parameter space. In our case, we translate the CMS constraints on ($m_{Z^{\prime}},g^{\prime}$) to our model, assuming order one $Z^\prime-\bar{\chi}-\chi$ coupling, $g_\chi=1$.
	
	\subsection{Neutrino trident}
	Neutrino trident production offers an additional and largely model--independent test of muonphilic gauge interactions. In the gauged $U(1)_{L_\mu-L_\tau}$ framework, the new $Z'$ boson couples to $\mu$ and $\nu_\mu$ (and with opposite sign to $\tau$ and $\nu_\tau$), modifying the cross section for $\nu_\mu N \to \nu_\mu N\,\mu^+\mu^-$ through interference with the SM $W/Z$ exchange. Comparing the predicted rate to the CCFR measurement of the trident cross section~\cite{CCFR:1991lpl} constrains the effective combination ${g^{\prime}}^2/m_{Z'}^2$, and can be expressed approximately as a lower bound on $m_{Z'}/g^{\prime}$ of a few hundred~GeV for $m_{Z'}\gtrsim \text{GeV}$. We directly use the constraints calculated in Ref.~\cite{Altmannshofer:2016jzy}.

	\subsection{Muon $g-2$}
	In both of our $U(1)'$ UV completions the $Z'$ couples \textit{vectorially} to muons, so the
	one-loop contribution to the muon anomalous magnetic moment depends only on
	$(g',m_{Z'})$ and is independent of whether the dark-sector current is vector or
	axial. In the heavy-mediator limit $m_{Z'}\gg m_\mu$, the $Z'$ contribution is well
	approximated as (see, for example, Ref.~\cite{Heeck:2011wj} for a review),
	\begin{equation}
	\Delta a_\mu \simeq (g'^2/12\pi^2)\,(m_\mu^2/m_{Z'}^2).
	\label{eq:amu_Zp}
	\end{equation}
	
	For the experimental value we use the final Fermilab Muon $g{-}2$ average,
	$a_\mu^{\rm exp}=116\,592\,0715(145)\times 10^{-12}$~\cite{Muong-2:2025xyk},
	and for the Standard Model prediction the 2025-Theory-Initiative-update,
	$a_\mu^{\rm SM}=116\,592\,033(62)\times 10^{-11}$~\cite{Aliberti:2025beg}.
	This gives
	\begin{equation}
	\Delta a_\mu^{\rm obs}=a_\mu^{\rm exp}-a_\mu^{\rm SM}
	=(38.5\pm 63.7)\times 10^{-11},
	\label{eq:amu_obs_2025}
	\end{equation}
	where the uncertainty is the quadrature combination of experimental and theory
	errors. Since $\Delta a_\mu(Z')>0$ for a purely vector muon coupling, we impose
	the one--sided $2\sigma$ upper bound:
	\begin{equation}
	\Delta a_\mu(Z') \le \Delta a_\mu^{\max}\equiv \Delta a_\mu^{\rm obs}+2\sigma_{\rm tot}
	\simeq 1.7\times 10^{-9},
	\label{eq:amu_deltamax}
	\end{equation}
	which translates, using Eq.~\eqref{eq:amu_Zp}, into
	\begin{equation}
	g' \le \frac{m_{Z'}}{200~\mathrm{GeV}}\,
	\sqrt{\frac{\Delta a_\mu^{\max}}{236\times 10^{-11}}}\,.
	\label{eq:gprime_bound_from_gm2}
	\end{equation}

	\section{Constraints and muon collider reach}
	\label{sec:global_bounds}
	
	\subsection{EFT interactions}
	\label{subsec:EFT}
\begin{figure}
		\centering
		\includegraphics[width=0.49\linewidth]{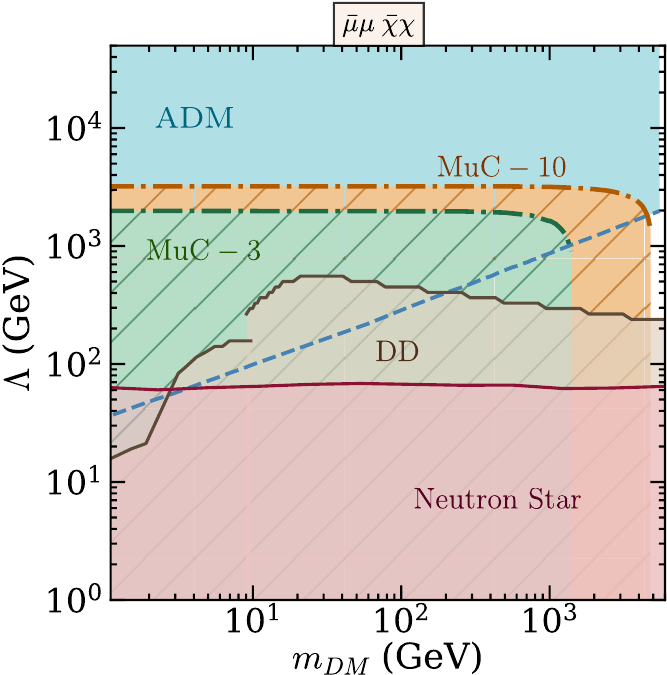}
		\includegraphics[width=0.49\linewidth]{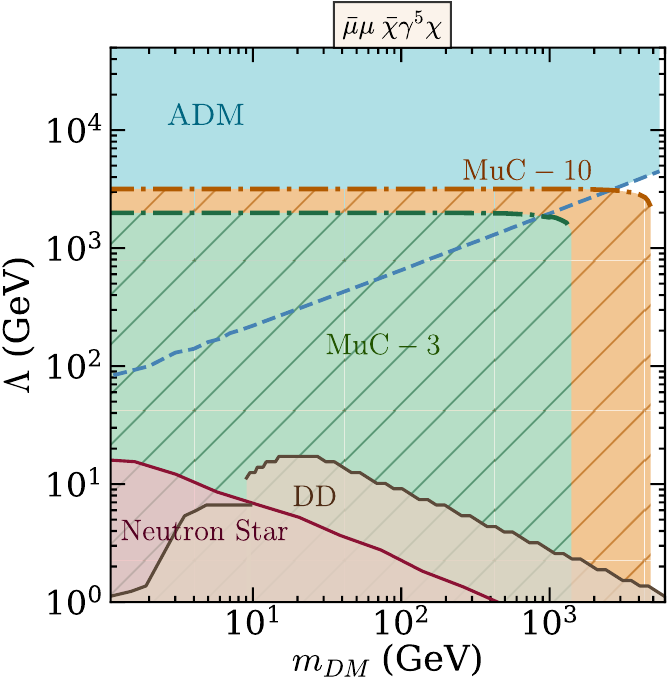}
		\caption{\small Limits on $\Lambda$ for $\mathcal{O}_{ss}\equiv \bar{\mu}\mu\,\bar{\chi}\chi$ (left) and $\mathcal{O}_{sp}\equiv\bar{\mu}\mu\,\bar{\chi}\gamma^5\chi$ (right) type interaction of DM with muons from different experimental observations and the ADM condition of Eq.~\ref{eq:constraint}, labeled on the respective regions with darker shades. The striped green and yellow region corresponds to the muon collider (MuC) 3$\sigma$ reach for $\sqrt{s}=3$ and 10 TeV respectively, with $\mathcal{L}=1~\text{ab}^{-1}$. DD corresponds to limits from DM-nucleon scattering experiments~\cite{PandaX:2022xqx, PandaX:2023xgl, LZ:2024zvo, PICO:2019vsc}. The experimental exclusions/reach extends up to the bottom of the plots and overlapping regions are implicit. See Sec.~\ref{subsec:EFT} for a discussion on the region  of validity of the EFT description. The neutron star sensitivity regions are as discussed in Sec.~\ref{sec:NS}.}
		\label{fig:EFT_exclusion1}
	\end{figure}

\begin{figure}
		\centering
		\includegraphics[width=0.49\linewidth]{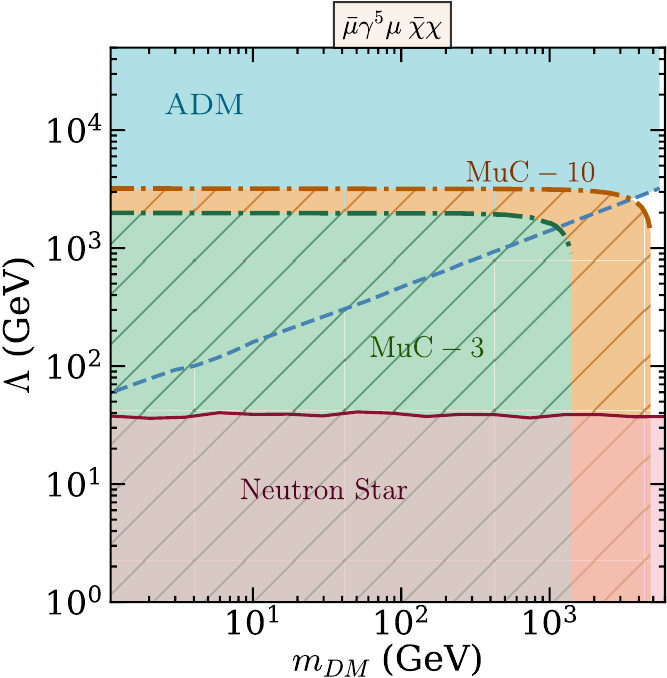}
		\includegraphics[width=0.49\linewidth]{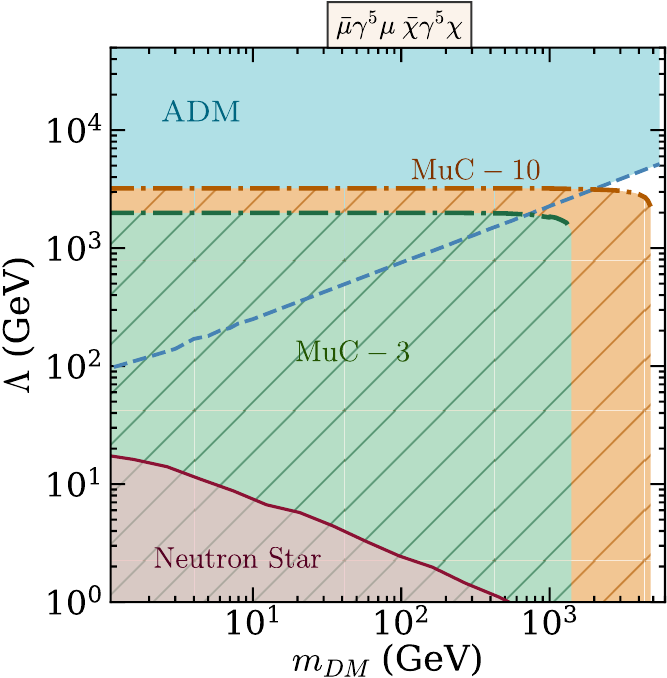}
		\caption{\small Similar to Figure~\ref{fig:EFT_exclusion1} for $\mathcal{O}_{ps}\equiv \bar\mu\gamma^5\mu\bar{\chi}\chi$ (left) and $\mathcal{O}_{pp}\equiv\bar{\mu}\gamma^5\mu\,\bar{\chi}\gamma^5\chi$ (right) type interaction.}
		\label{fig:EFT_exclusion2}
	\end{figure}

\begin{figure}
		\centering
		\includegraphics[width=0.49\linewidth]{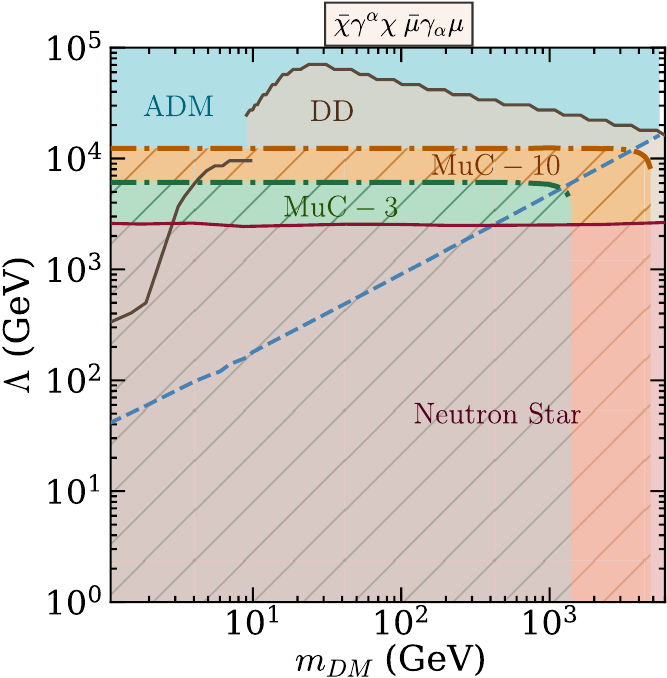}
		\includegraphics[width=0.49\linewidth]{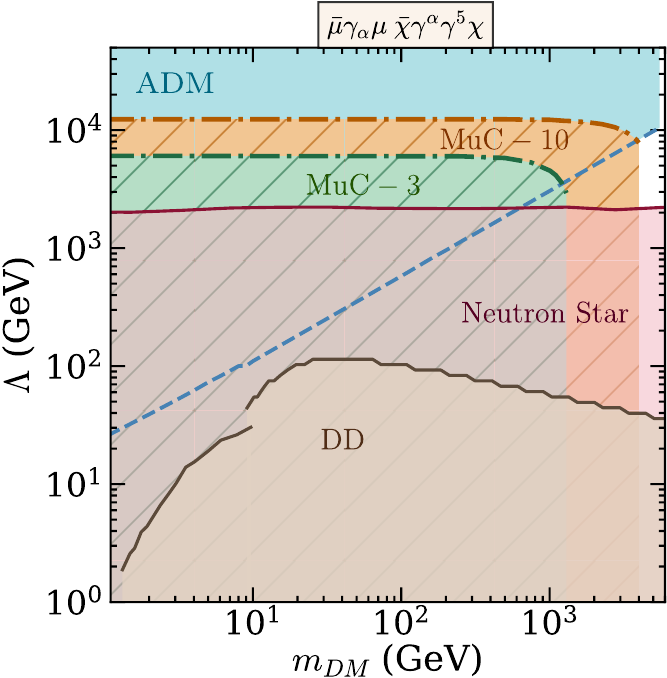}
		\caption{\small Similar to Figure~\ref{fig:EFT_exclusion1} for $\mathcal{O}_{vv}\equiv \bar\mu\gamma_\alpha\mu\,\bar{\chi}\gamma^\alpha\chi$ (left) and $\mathcal{O}_{va}\equiv\bar\mu\gamma_\alpha\mu\,\bar{\chi}\gamma^\alpha\gamma^5\chi$ (right) type interaction.}
		\label{fig:EFT_exclusion3}
	\end{figure}

\begin{figure}
		\centering
		\includegraphics[width=0.49\linewidth]{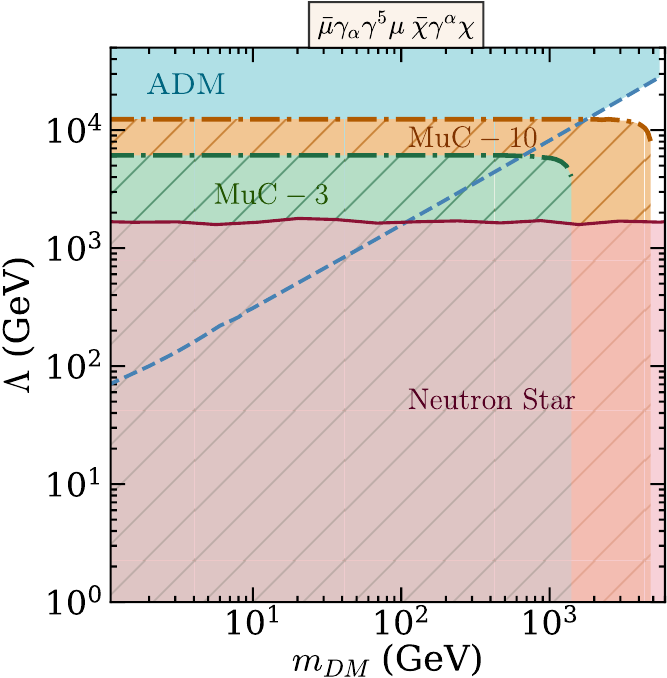}
		\includegraphics[width=0.49\linewidth]{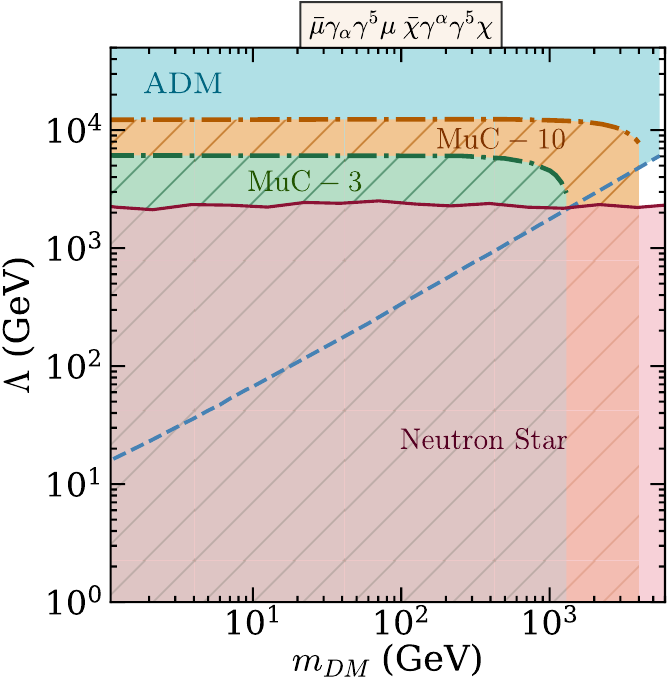}
		\caption{\small Similar to Figure~\ref{fig:EFT_exclusion1} for $\mathcal{O}_{av}\equiv \bar\mu\gamma_\alpha\gamma^5\mu\,\bar{\chi}\gamma^\alpha\chi$ (left) and $\mathcal{O}_{aa}\equiv\bar\mu\gamma_\alpha\gamma^5\mu\,\bar{\chi}\gamma^\alpha\gamma^5\chi$ (right) type interaction.}
		\label{fig:EFT_exclusion4}
	\end{figure}

\begin{figure}
		\centering
		\includegraphics[width=0.49\linewidth]{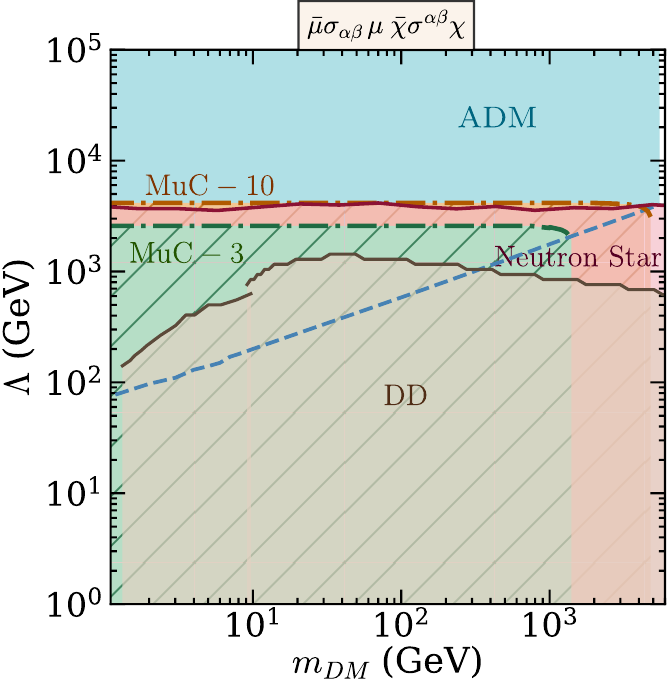}
		\includegraphics[width=0.49\linewidth]{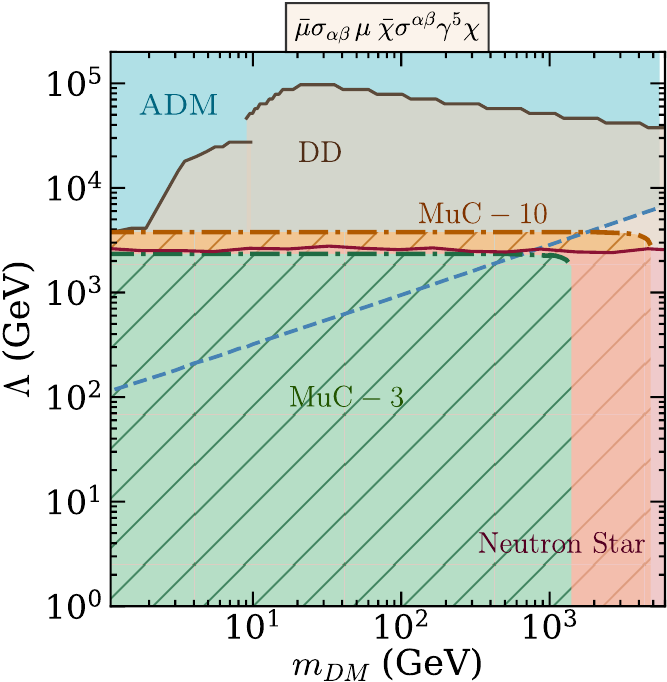}
		\caption{\small Similar to Figure~\ref{fig:EFT_exclusion1} for $\mathcal{O}_{tt}\equiv \bar\mu\sigma_{\alpha\beta}\mu\,\bar{\chi}\sigma^{\alpha\beta}\chi$ (left) and $\mathcal{O}_{pt}\equiv\bar\mu\sigma_{\alpha\beta}\mu\,\bar{\chi}\sigma^{\alpha\beta}\gamma^5\chi$ (right) type interaction.}
		\label{fig:EFT_exclusion5}
	\end{figure}

	Figures~\ref{fig:EFT_exclusion1}–\ref{fig:EFT_exclusion5} show the $3\sigma$ sensitivity to various EFT interactions of muonphilic ADM (see Table~\ref{tab:operators_muon}) in the $\Lambda$--$m_{\chi}$ plane at a future muon collider, for two center-of-mass energies, $\sqrt{s}=3$ and $10$ TeV, and an integrated luminosity $\mathcal{L}=1~\text{ab}^{-1}$.\footnote{We have checked that increasing the integrated luminosity to 10 ab$^{-1}$ increases the sensitivity by $\sim 20-30\%$. However, this does not change our general conclusions.} 
	
	A partial-wave unitarity estimate also clarifies the regime of validity of the four-fermion EFT description used for $\mu^+\mu^- \to \chi\bar{\chi}$ at a future muon collider. In the light $\chi$ regime, the EFT description is not reliable for $\Lambda \lesssim 0.4~\text{TeV}$ at $\sqrt{s}=3~\text{TeV}$ and for $\Lambda \lesssim 1.4~\text{TeV}$ at $\sqrt{s}=10~\text{TeV}$, up to $\mathcal{O}(1)$ factors depending on the Lorentz structure of the operator.\footnote{We estimate that the vector and axial vector unitarity bounds are about a factor of two stronger, while the bounds for tensors are slightly weaker.} For example, for the scalar operator $\mathcal{O}_{ss}=(\bar{\mu}\mu)(\bar{\chi}\chi)/\Lambda^2$, one obtains
	\begin{align}
	\Lambda_{\rm unit}^{(ss)}(s,m_\chi) \gtrsim \sqrt{\frac{s\,\beta_f^{3/2}}{16\pi}},
	\qquad
	\beta_f = \sqrt{1-\frac{4m_\chi^2}{s}} \, .
	\end{align}
	This shows that, for the muon collider probe, the EFT interpretation of the projected sensitivities in Figs.~\ref{fig:EFT_exclusion1}--\ref{fig:EFT_exclusion5} should be understood as reliable only in the region where the inferred values of $\Lambda$ remain above the corresponding unitarity thresholds. We also note that as $m_\chi$ approaches the kinematic threshold, the bound weakens because of the $\beta_f$ suppression, following the same qualitative trend as the reduction in the muon-collider sensitivity in that region.
	
	The blue regions in these figures correspond to parameter space that violates the ADM criterion of Eq.~\ref{eq:constraint}. We contrast this sensitivity with constraints from DD experiments and from studies of DM capture in neutron stars. 
	
	The DD bounds exclude ADM up to scales of almost $10$ TeV for the operators $\op_{pt}$ and $\op_{vv}$, and is slightly weaker for $\op_{ss}$, at the level of $\mathcal{O}(100)$ GeV. By contrast, they are much weaker for the spin-dependent DD interactions
	$\mathcal{O}_{tt}$ and $\mathcal{O}_{va}$, and for operators
	with pseudoscalar or axial-vector interactions in the SM muon current
	($\mathcal{O}_{ps}$, $\mathcal{O}_{pp}$, $\mathcal{O}_{av}$,
	$\mathcal{O}_{aa}$), the DD constraints are essentially absent. For this latter set of
	operators, neutron-star sensitivities are strong for $\mathcal{O}_{va}$,
	$\mathcal{O}_{aa}$, and $\mathcal{O}_{tt}$, potentially probing the ADM scenario
	up to scales of $\mathcal{O}(100\text{--}1000)~\text{GeV}$. However, for the pseudo-scalar operators $\op_{sp}$, $\op_{ps}$ and $\op_{pp}$, only the muon collider is sensitive over the full DM mass range up to several TeV; even for $\op_{ss}$, $\op_{va}$ and $\op_{av}$, a region of parameter space at ADM masses above a few hundred GeV remains accessible only to a muon collider. 
	
	We have also checked that using chiral muonphilic portals, namely purely right- or left-handed muons with either vector-like or axial-vector-like $\chi$, in place of the individual contact operators in Table~\ref{tab:operators_muon} does not lead to any qualitative change in our results: the vector-like DM cases closely follow Fig.~\ref{fig:EFT_exclusion3} (left), while the axial-vector cases closely follow Fig.~\ref{fig:EFT_exclusion3} (right). This is because the relevant ADM, DD, NS and muon-collider observables depend only on quadratic combinations of the WCs, since the interference terms for the above operator combinations are zero. The corresponding exclusion curves are thus only $\mathcal{O}(1)$ rescalings of the single-operator bounds.
	
	Note that Figs.~\ref{fig:EFT_exclusion1}-\ref{fig:EFT_exclusion5} display the scale of new physics $\Lambda$ down to as low as a few GeV. While one may, in general, have legitimate concerns about the existence of phenomenologically acceptable UV completions for such low $\Lambda$ scales, it has been shown that in some cases there are in fact viable completions~\cite{Guchait:2020wqn}. We therefore consider it legitimate to countenance low values of $\Lambda$ for all of the EFT operators.
	
	Finally, observe that constraints from dimuon resonance searches, neutrino trident production, and muon $g-2$ primarily probe the mediator-muon sector (and for the trident process, the associated $\nu_{\mu}$ coupling). In a bottom-up EFT where the WC is treated as free, these bounds cannot be mapped unambiguously onto the EFT parameter space without specifying the UV completion (mediator mass, separate muon and DM couplings, etc.). We therefore apply them only to the explicit $Z^{\prime}$ benchmark models.

	\subsection{UV-models}
\begin{figure}
		\centering
		\includegraphics[width=0.32\linewidth]{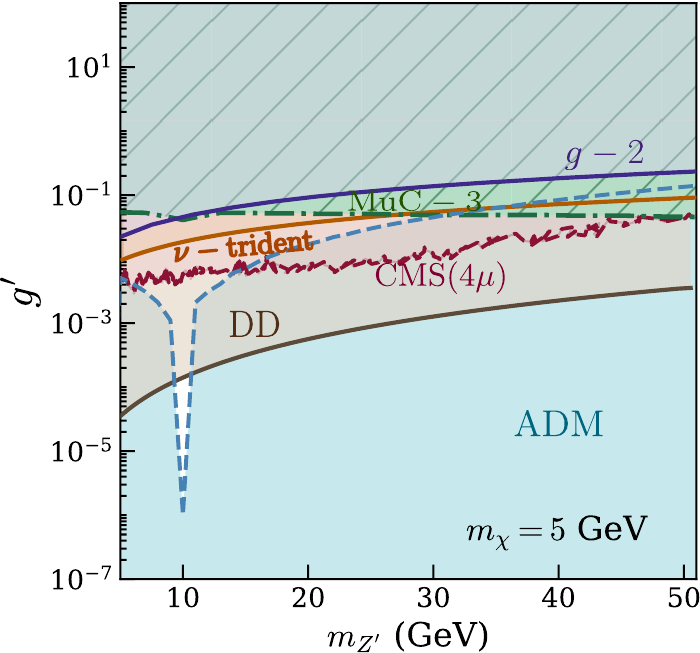}
		\includegraphics[width=0.32\linewidth]{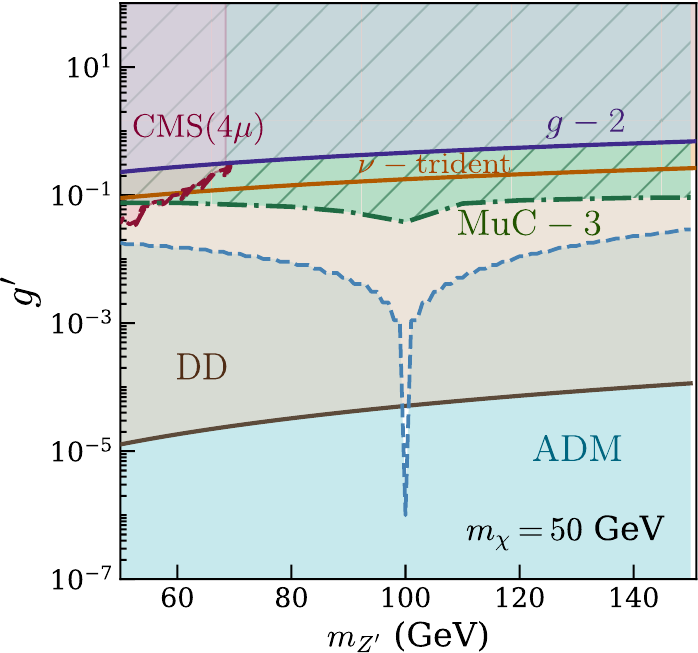}
		\includegraphics[width=0.33\linewidth]{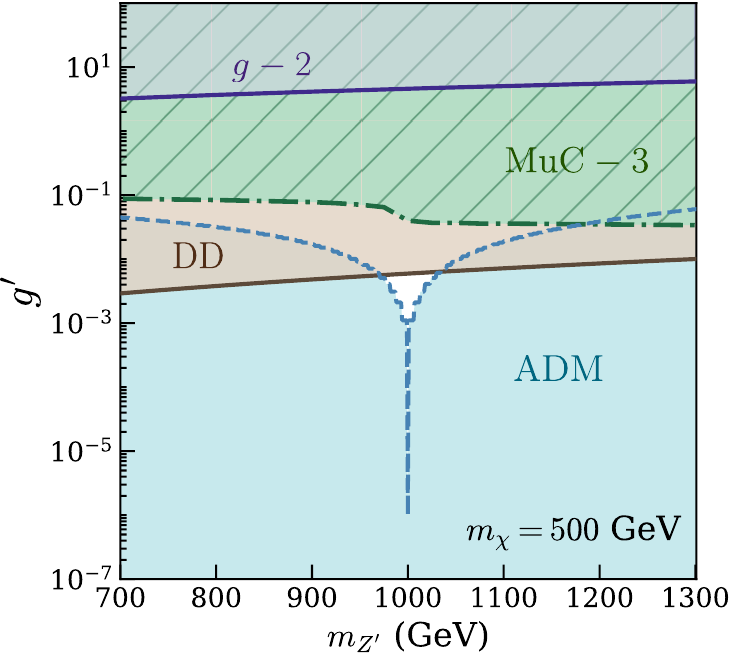}
		\caption{\small Constraints on $Z^\prime$–muon interactions in the vector-$L_\mu-L_\tau$ model, shown as constraints on the gauge coupling $g^{\prime}$ from various experimental observations, for three ADM masses: $\rm m_\chi=5$ GeV (left), $\rm m_\chi=50$ GeV (center), and $\rm m_\chi=500$ GeV (right). All exclusion regions extend to the top edge of the plot, and overlapping regions are implied. The projected $3\sigma$ sensitivity from the muon collider (MuC) is for $\sqrt{s}=3$ TeV, $\mathcal{L}=3~\text{ab}^{-1}$, and $g_\chi$=1. The ADM-disfavored region is shown in blue. CMS and $\nu$-trident bounds are taken from \cite{CMS:2018yxg}, and the muon $g-2$ bound is treated as explained in the main text. In all cases, the allowed parameter region is very small and fine-tuned to the resonant $Z'$ regime. The 3 TeV MuC sensitivity is found to be below that of DD experiments.}
		\label{fig:UV_exclusion}
	\end{figure}

\begin{figure}
		\centering
		\includegraphics[width=0.33\linewidth]{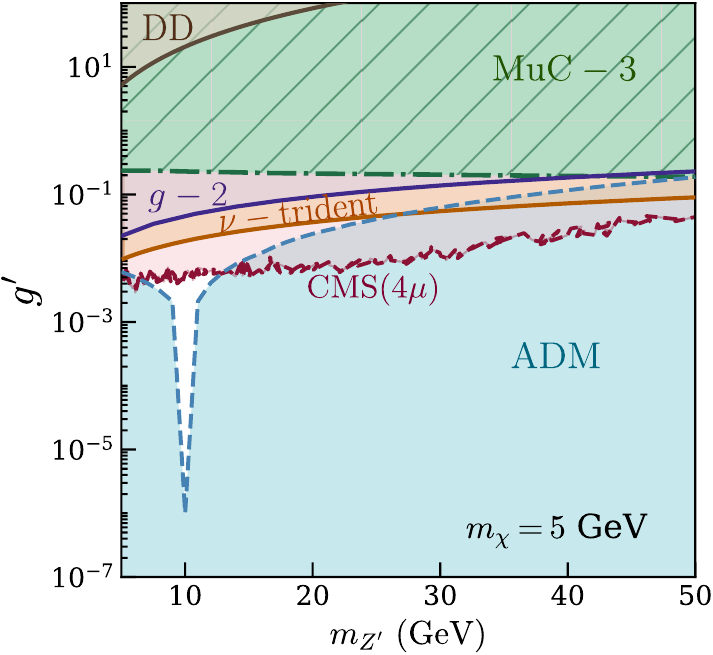}
		\includegraphics[width=0.32\linewidth]{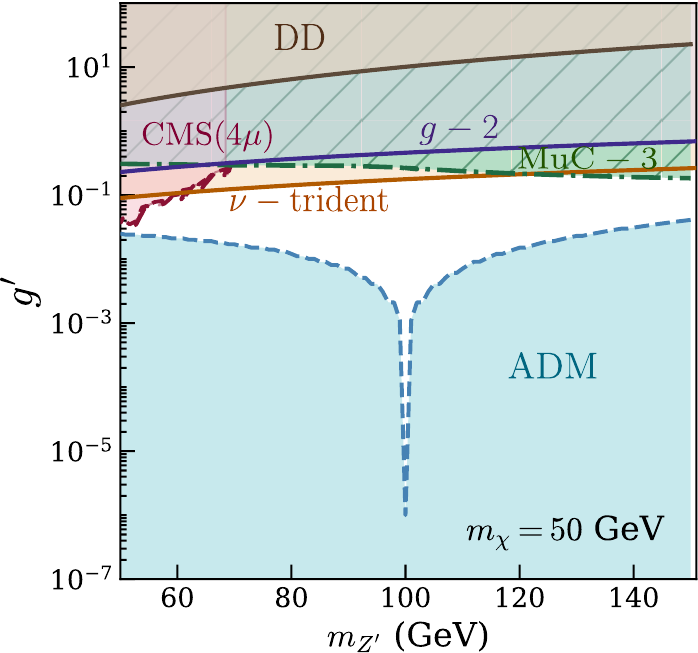}
		\includegraphics[width=0.33\linewidth]{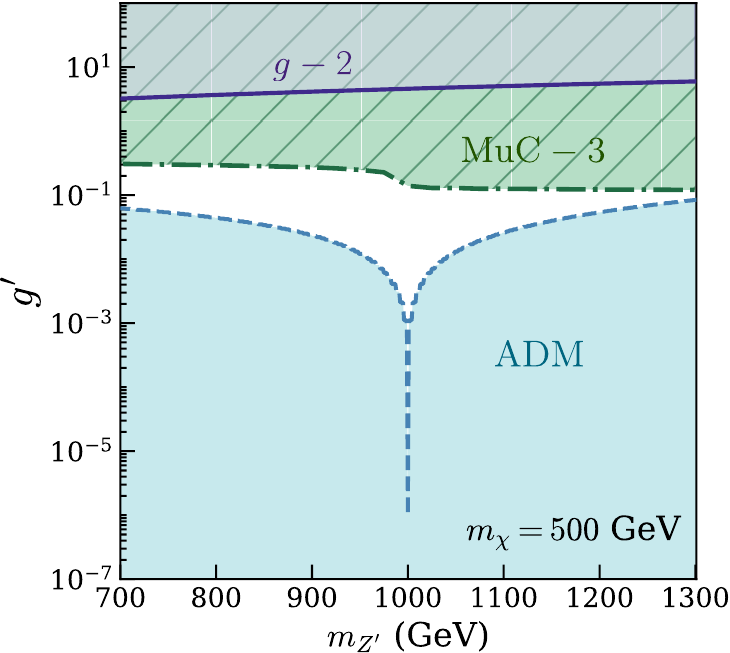}
		\caption{\small Same as Fig.~\ref{fig:UV_exclusion} for the axial-$\lmultau$ model. Note that in this case the allowed parameter region is larger, and a muon collider has the sensitivity to probe additional parameter space for high DM masses such as the 500 GeV depicted in the rightmost plot.}
		\label{fig:UV_exclusion_axial}
	\end{figure}

	In Figs.~\ref{fig:UV_exclusion} and \ref{fig:UV_exclusion_axial} we show the constraints on the parameter spaces of the UV models defined in Secs.~\ref{subsec:vectormodel} and \ref{subsec:axialvectormodel}. For these figures, we choose three benchmark ADM masses $m_{\chi}=5$ GeV, $50$ GeV and $500$ GeV, and the $Z^\prime-\chi-\chi$ coupling $g_\chi=1$ in general. Additionally, for the axial $\lmultau$ model, we choose charge $q=1$ and the mixing angle $\theta=10^{-2}$, implying a small mixing scenario. As we can see in Fig.~\ref{fig:UV_exclusion}, the DD bounds almost entirely exclude the possibility of ADM in the $\mathcal{O}(1-100)$ GeV mass range for the vector $\lmultau$, except a narrow patch on the $Z^\prime$-resonance, $m_{Z^\prime}\simeq 2m_{\chi}$. For the axial $\lmultau$ model, on the other hand, there are viable regions of parameter space allowed by all the current experiments for an ADM of mass $m_{\chi}=50$ GeV, as we see in Fig.~\ref{fig:UV_exclusion_axial}. The bounds get weaker once we move to higher ADM masses.
	
	For both UV models, we conclude that a muon collider cannot beat the sensitivity of the other experiments in the 1-100 GeV mass range favoured by ADM solutions to the coincidence problem. However, for general ADM with higher masses (e.g.\ 500 GeV, as per the rightmost plot of Fig.~\ref{fig:UV_exclusion_axial}), a muon collider would have the sensitivity to probe additional parameter space of the axial UV model.

	\section{Conclusion}
	\label{sec:conclusion}
	
	We have analysed muonphilic portals for fermionic asymmetric dark matter through both effective operators and two UV models based on gauged $L_\mu - L_\tau$, one featuring vector coupling to the DM and the other axial coupling. We derived constraints from (i) the ADM criterion that at least $99\%$ of the relic DM density is asymmetric, and (ii) the latest direct detection results. Furthermore, we indicated prospective sensitivities from neutron star heating for the EFT interactions, and constraints from neutrino trident production, CMS $pp \to 4\mu$ searches and $g-2$ of the muon for the UV cases, to obtain a global picture. We then computed the projected sensitivities of future 3 and 10 TeV muon colliders to probe the viable regions of parameter space with 1 ab$^{-1}$ of data. The results, as depicted in Figs.~\ref{fig:EFT_exclusion1}–\ref{fig:UV_exclusion_axial}, can be summarised as follows:
	
	For the few-GeV DM mass range favoured by ADM solutions to the $\Omega_b \sim \Omega_\text{DM}/5$ coincidence problem, the operators $\mathcal{O}_{ss,sp,ps,pp,va,av,aa}$ have viable parameter spaces that have potentially observable implications for neutron star heating. Note that the region is very small for $\mathcal{O}_{ss}$. The operators $\mathcal{O}_{vv,pt}$ have no viable parameter space in the entire mass range. For the regions allowed by the ADM condition in the few-GeV DM mass range, the effective operator analysis is unreliable for computing muon collider sensitivities because the unitarity violation bound is exceeded for the muon collider energy scales. However, the muon collider sensitivities may still be applicable for a generic DM model with these EFT interactions.

	For the mass range above $\text{few} \times 100$ GeV, as relevant for general ADM probably unconnected with the coincidence problem, the effective operator portals $\mathcal{O}_{ss,sp,ps,pp,va,av,aa}$ have viable parameter space ranges that can be probed by 3 and 10 TeV muon colliders. The operators $\mathcal{O}_{vv}$ and $\mathcal{O}_{pt}$ are ruled out by direct-detection constraints, while $\mathcal{O}_{tt}$ can be probed with comparable sensitivity by a $10~\text{TeV}$ muon collider and through neutron-star heating effects.
	
	The vector gauged $L_\mu - L_\tau$ model is severely constrained by existing bounds and the ADM criterion, permitting only fine-tuned regions near the $m_{Z'} \simeq 2m_\chi$ resonance regime. Those small regions, however, will be inaccessible to any plausible muon collider facility. 
	
	The axial gauged $L_\mu - L_\tau$ model, on the other hand, has a significantly larger region of allowed parameter space, which increases in size as the DM mass is made larger. The parameter space for relatively large DM masses such as 500 GeV can be further explored by a muon collider.
	
	\section*{Acknowledgements}
	
	We thank Nicole Bell, Giorgio Busoni, and Basudeb Dasgupta for helpful discussions, and Giorgio Busoni for useful feedback on a draft of this paper. This work was supported
	in part by the Australian Research Council through the ARC Centre of Excellence for Dark Matter Particle Physics, CE200100008. The work of AR is supported by the ARC-DP230101142.
	
	\newpage
	\appendix
	\section{Appendix: Catalogue of annihilation rates $\langle\sigma v\rangle$}
	\label{app:annihilation}
	\paragraph{I.\ $\langle\sigma v\rangle$ for EFT interactions:}
	The following are the expressions for the annihilation rates as a function of ADM mass $m_\chi$, muon mass $m_{\mu}$, EFT-cut-off scale $\Lambda$, and the relative velocity $v$ of the annihilating DM for the operators in Table \ref{tab:operators_muon},
	
	\begin{align}
	\mathcal{O}_{ss}: \langle\sigma^{}v\rangle &=\frac{3m_{\chi}^2\,v_h^2}{8\pi\Lambda^6}\left(1-\frac{m_\mu^2}{m_{\chi}^2}\right)^{3/2}v^2,\\
	\mathcal{O}_{pp}: \langle\sigma^{}v\rangle &=\frac{3m_{\chi}^2\,v_h^2}{\pi\Lambda^6}\left(1-\frac{m_\mu^2}{m_{\chi}^2}\right)^{1/2} \left[1+\frac{v^2}{8}\left(\frac{2m_{\chi}^2-m_\mu^2}{m_{\chi}^2-m_\mu^2}\right)\right],\\
	\mathcal{O}_{ps}: \langle\sigma^{}v\rangle &=\frac{3m_{\chi}^2\,v_h^2}{4\pi\Lambda^6}\left(1-\frac{m_\mu^2}{m_{\chi}^2}\right)^{1/2}v^2,\\
	\mathcal{O}_{sp}: \langle\sigma^{}v\rangle &=\frac{3m_{\chi}^2\,v_h^2}{4\pi\Lambda^6}\left(1-\frac{m_\mu^2}{m_{\chi}^2}\right)^{3/2}\left[1+v^2\left(\frac{2m_\mu^2-m_{\chi}^2}{m_{\chi}^2-m_\mu^2}\right)\right],\\	
	\mathcal{O}_{vv}: \langle\sigma^{}v\rangle &=\frac{3m_{\chi}^2}{2\pi\Lambda^4}\left(1-\frac{m_\mu^2}{m_{\chi}^2}\right)^{1/2}\left[\left(2+\frac{m_\mu^2}{m_{\chi}^2}\right)
	+v^2\left(\frac{8m_{\chi}^4-4m_\mu^2m_{\chi}^2+5m_\mu^4}{24m_{\chi}^2(m_{\chi}^2-m_\mu^2)}\right)\right],\\
	\mathcal{O}_{aa}: \langle\sigma^{}v\rangle &=\frac{3m_{\chi}^2}{2\pi\Lambda^4}\left(1-\frac{m_\mu^2}{m_{\chi}^2}\right)^{1/2}\left[\frac{m_\mu^2}{m_{\chi}^2}
	+v^2\left(\frac{8m_{\chi}^4-22m_\mu^2 m_{\chi}^2+17m_\mu^4}{24m_{\chi}^2(m_{\chi}^2-m_\mu^2)}\right)\right],\\
	\mathcal{O}_{av}: \langle\sigma^{}v\rangle &=\frac{3m_{\chi}^2}{\pi\Lambda^4}\left(1-\frac{m_\mu^2}{m_{\chi}^2}\right)^{1/2}\left[\left(1-\frac{m_\mu^2}{m_{\chi}^2}\right)+\frac{v^2}{24}\left(4+5\frac{m_\mu^2}{m_{\chi}^2}\right)\right],\\
	\mathcal{O}_{va}: \langle\sigma^{}v\rangle &=\frac{m_{\chi}^2}{4\pi\Lambda^4}\left(1-\frac{m_\mu^2}{m_{\chi}^2}\right)^{1/2}\left(2+\frac{m_\mu^2}{m_{\chi}^2}\right)v^2,\\
	\mathcal{O}_{tt}:  \langle\sigma^{}v\rangle &=\frac{3m_{\chi}^2\,v_h^2}{\pi\Lambda^6}\left(1-\frac{m_\mu^2}{m_{\chi}^2}\right)^{1/2}\left[\left(1+2\frac{m_\mu^2}{m_{\chi}^2}\right)
	+v^2\left(\frac{4m_{\chi}^4-11m_\mu^2 m_{\chi}^2+16m_\mu^4}{24m_{\chi}^2(m_{\chi}^2-m_\mu^2)}\right)\right],\\
	\mathcal{O}_{pt}: \langle\sigma^{}v\rangle &=\frac{6m_{\chi}^2\,v_h^2}{\pi\Lambda^6}\left(1-\frac{m_\mu^2}{m_{\chi}^2}\right)^{1/2}\left[\left(1-\frac{m_\mu^2}{m_{\chi}^2}\right)+\frac{v^2}{24}\left(4+11\frac{m_\mu^2}{m_{\chi}^2}\right)\right].
	\end{align}
	
	\vspace{10mm}
	
	\paragraph{II.\ $\langle\sigma v\rangle$ for the vector $U(1)_{\lmultau}$ model:}    
	
	\begin{align}
	\langle\sigma v\rangle_{\chi\bar\chi\to Z'Z'} &= \frac{g_\chi^4}{16\pi m_\chi^2} 
	\left( 1 - \frac{m_{Z'}^2}{m_\chi^2} \right)^{3/2}
	\left( 1 - \frac{m_{Z'}^2}{2m_\chi^2} \right)^{-2} \ ,
	\label{eq:sigv_ZpZp}\\[6pt]
	\langle\sigma v\rangle_{\chi\bar\chi\to \nu_\ell \bar\nu_\ell} &= 
	\frac{g_\chi^2 {g^{\prime}}^2 m_\chi^2}{2\pi\big[(4m_\chi^2 - m_{Z'}^2)^2 + m_{Z'}^2 \Gamma_{Z'}^2\big]} \ ,
	\label{eq:sigv_nunu}\\[6pt]
	\langle\sigma v\rangle_{\chi\bar\chi\to \ell^+\ell^-} &=
	\frac{g_\chi^2 {g^{\prime}}^2 m_\chi^2}{\pi\big[(4m_\chi^2 - m_{Z'}^2)^2 + m_{Z'}^2 \Gamma_{Z'}^2\big]}
	\left( 1 + \frac{m_{\ell}^2}{2m_\chi^2} \right)
	\sqrt{1 - \frac{m_{\ell}^2}{m_\chi^2}} \ ,
	\label{eq:sigv_ll}
	\end{align}
	where $\Gamma_{Z'}$ is the total $Z'$ width, including all kinematically allowed decays, e.g. into charged leptons, neutrinos and $\chi\bar\chi$. Equations \eqref{eq:sigv_nunu} and \eqref{eq:sigv_ll} are written per neutrino/charged-lepton species. For the $L_\mu-L_\tau$ charge assignment the invisible annihilation receives contributions from both $\nu_\mu$ and $\nu_\tau$. The $Z'Z'$ channel (Eq.~\ref{eq:sigv_ZpZp}) typically dominates once open and efficiently depletes the symmetric component.
	
	\vspace{10mm}
	
	\paragraph{III.\ $\langle\sigma v\rangle$ for the axial $U(1)_{\lmultau}$ model:}
	
	Similar to the vector $U(1)_{L_\mu-L_\tau}$ case, the non-relativistic rates are
	\begin{align}
	\langle \sigma v \rangle_{X_1 \bar{X}_1 \to Z' Z'} &=
	\frac{g_{\text{ax}}^4}{16 \pi m_1^2}
	\left( 1 - \frac{m_{Z'}^2}{m_1^2} \right)^{3/2}
	\left( 1 - \frac{m_{Z'}^2}{2 m_1^2} \right)^{-2}, \\
	\langle \sigma v \rangle_{X_1 \bar{X}_1 \to \nu_\ell \bar{\nu}_\ell} &=
	\frac{g_{\text{ax}}^2 g'^2 m_1^2}{2 \pi (4m_1^2 - m_{Z'}^2)^2 + m_{Z'}^2 \Gamma_{Z'}^2} \quad (\ell = \mu, \tau), \\
	\langle \sigma v \rangle_{X_1 \bar{X}_1 \to \ell^+ \ell^-} &=
	\frac{g_{\text{ax}}^2 g'^2}{2 \pi}
	\frac{1}{(4m_1^2 - m_{Z'}^2)^2 + m_{Z'}^2 \Gamma_{Z'}^2}
	\left[
	\frac{m_\ell^2}{2}
	+ \frac{v^2}{6} m_1^2 \left( 1 + \frac{m_\ell^2}{2 m_1^2} \right)
	\right],
	\end{align}
	where, $g_{\text{ax}}$ is the gauge coupling for the axial-$\lmultau$ model (see Eq.~\ref{eq:gax}), \text{and} $m_1$ is the mass of the ADM particle $X_1$. 
	
	\bibliographystyle{JHEP}
	\bibliography{adm.bib}

@article{Cohen:2009fz,
    author = "Cohen, Timothy and Zurek, Kathryn M.",
    title = "{Leptophilic Dark Matter from the Lepton Asymmetry}",
    eprint = "0909.2035",
    archivePrefix = "arXiv",
    primaryClass = "hep-ph",
    reportNumber = "FERMILAB-PUB-09-426-T, MCTP-09-45",
    doi = "10.1103/PhysRevLett.104.101301",
    journal = "Phys. Rev. Lett.",
    volume = "104",
    pages = "101301",
    year = "2010"
}

@article{Blennow:2010qp,
    author = "Blennow, Mattias and Dasgupta, Basudeb and Fernandez-Martinez, Enrique and Rius, Nuria",
    title = "{Aidnogenesis via Leptogenesis and Dark Sphalerons}",
    eprint = "1009.3159",
    archivePrefix = "arXiv",
    primaryClass = "hep-ph",
    reportNumber = "MPP-2010-125, IFIC-10-32, FTUV-10-0909",
    doi = "10.1007/JHEP03(2011)014",
    journal = "JHEP",
    volume = "03",
    pages = "014",
    year = "2011"
}

@article{Foot:1990mn,
    author = "Foot, Robert",
    title = "{New physics from electric charge quantization?}",
    reportNumber = "MAD/TH/90-14",
    doi = "10.1142/S0217732391000543",
    journal = "Mod. Phys. Lett. A",
    volume = "6",
    pages = "527--530",
    year = "1991"
}

@article{He:1990pn,
    author = "He, X.-G. and Joshi, Girish C. and Lew, H. and Volkas, R. R.",
    title = "{New $Z'$ phenomenology}",
    reportNumber = "UM-P-90/42, OZ-P-90/16",
    doi = "10.1103/PhysRevD.43.R22",
    journal = "Phys. Rev. D",
    volume = "43",
    pages = "22--24",
    year = "1991"
}

@article{He:1991qd,
    author = "He, Xiao-Gang and Joshi, Girish C. and Lew, H. and Volkas, R. R.",
    title = "{Simplest $Z'$ model}",
    reportNumber = "CERN-TH-6084-91, UM-P-91-32, OZ-91-07",
    doi = "10.1103/PhysRevD.44.2118",
    journal = "Phys. Rev. D",
    volume = "44",
    pages = "2118--2132",
    year = "1991"
}

@article{Foot:1994vd,
    author = "Foot, Robert and He, X.-G. and Lew, H. and Volkas, R. R.",
    title = "{Model for a light $Z'$ boson}",
    eprint = "hep-ph/9401250",
    archivePrefix = "arXiv",
    reportNumber = "OITS-532, UM-P-93-115, OZ-93-26, IP-ASTP-32",
    doi = "10.1103/PhysRevD.50.4571",
    journal = "Phys. Rev. D",
    volume = "50",
    pages = "4571--4580",
    year = "1994"
}

@article{Strigari:2009bq,
    author = "Strigari, Louis E.",
    title = "{Neutrino Coherent Scattering Rates at Direct Dark Matter Detectors}",
    eprint = "0903.3630",
    archivePrefix = "arXiv",
    primaryClass = "astro-ph.CO",
    doi = "10.1088/1367-2630/11/10/105011",
    journal = "New J. Phys.",
    volume = "11",
    pages = "105011",
    year = "2009"
}

@article{Baryakhtar:2017dbj,
    author = "Baryakhtar, Masha and Bramante, Joseph and Li, Shirley Weishi and Linden, Tim and Raj, Nirmal",
    title = "{Dark Kinetic Heating of Neutron Stars and An Infrared Window On WIMPs, SIMPs, and Pure Higgsinos}",
    eprint = "1704.01577",
    archivePrefix = "arXiv",
    primaryClass = "hep-ph",
    doi = "10.1103/PhysRevLett.119.131801",
    journal = "Phys. Rev. Lett.",
    volume = "119",
    number = "13",
    pages = "131801",
    year = "2017"
}

@article{Raj:2017wrv,
    author = "Raj, Nirmal and Tanedo, Philip and Yu, Hai-Bo",
    title = "{Neutron stars at the dark matter direct detection frontier}",
    eprint = "1707.09442",
    archivePrefix = "arXiv",
    primaryClass = "hep-ph",
    reportNumber = "UCR-TR-2017-FLIP-NCC-1701",
    doi = "10.1103/PhysRevD.97.043006",
    journal = "Phys. Rev. D",
    volume = "97",
    number = "4",
    pages = "043006",
    year = "2018"
}

@article{Billard:2013qya,
    author = "Billard, J. and Strigari, L. and Figueroa-Feliciano, E.",
    title = "{Implication of neutrino backgrounds on the reach of next generation dark matter direct detection experiments}",
    eprint = "1307.5458",
    archivePrefix = "arXiv",
    primaryClass = "hep-ph",
    doi = "10.1103/PhysRevD.89.023524",
    journal = "Phys. Rev. D",
    volume = "89",
    number = "2",
    pages = "023524",
    year = "2014"
}

@article{Monroe:2007xp,
    author = "Monroe, Jocelyn and Fisher, Peter",
    title = "{Neutrino Backgrounds to Dark Matter Searches}",
    eprint = "0706.3019",
    archivePrefix = "arXiv",
    primaryClass = "astro-ph",
    doi = "10.1103/PhysRevD.76.033007",
    journal = "Phys. Rev. D",
    volume = "76",
    pages = "033007",
    year = "2007"
}

@article{Zurek:2013wia,
    author = "Zurek, Kathryn M.",
    title = "{Asymmetric Dark Matter: Theories, Signatures, and Constraints}",
    eprint = "1308.0338",
    archivePrefix = "arXiv",
    primaryClass = "hep-ph",
    doi = "10.1016/j.physrep.2013.12.001",
    journal = "Phys. Rept.",
    volume = "537",
    pages = "91--121",
    year = "2014"
}

@article{Petraki:2013wwa,
    author = "Petraki, Kalliopi and Volkas, Raymond R.",
    title = "{Review of asymmetric dark matter}",
    eprint = "1305.4939",
    archivePrefix = "arXiv",
    primaryClass = "hep-ph",
    reportNumber = "NIKHEF-2013-016",
    doi = "10.1142/S0217751X13300287",
    journal = "Int. J. Mod. Phys. A",
    volume = "28",
    pages = "1330028",
    year = "2013"
}

@article{Iminniyaz:2011yp,
    author = "Iminniyaz, Hoernisa and Drees, Manuel and Chen, Xuelei",
    title = "{Relic Abundance of Asymmetric Dark Matter}",
    eprint = "1104.5548",
    archivePrefix = "arXiv",
    primaryClass = "hep-ph",
    doi = "10.1088/1475-7516/2011/07/003",
    journal = "JCAP",
    volume = "07",
    pages = "003",
    year = "2011"
}

@article{March-Russell:2012elz,
    author = "March-Russell, John and Unwin, James and West, Stephen M.",
    title = "{Closing in on Asymmetric Dark Matter I: Model independent limits for interactions with quarks}",
    eprint = "1203.4854",
    archivePrefix = "arXiv",
    primaryClass = "hep-ph",
    reportNumber = "OUTP-12-01P",
    doi = "10.1007/JHEP08(2012)029",
    journal = "JHEP",
    volume = "08",
    pages = "029",
    year = "2012"
}

@article{Arcadi:2021mag,
    author = "Arcadi, Giorgio and Djouadi, Abdelhak and Kado, Marumi",
    title = "{The Higgs-portal for dark matter: effective field theories versus concrete realizations}",
    eprint = "2101.02507",
    archivePrefix = "arXiv",
    primaryClass = "hep-ph",
    doi = "10.1140/epjc/s10052-021-09411-2",
    journal = "Eur. Phys. J. C",
    volume = "81",
    number = "7",
    pages = "653",
    year = "2021"
}

@article{PICO:2019vsc,
    author = "Amole, C. and others",
    collaboration = "PICO",
    title = "{Dark Matter Search Results from the Complete Exposure of the PICO-60 C$_3$F$_8$ Bubble Chamber}",
    eprint = "1902.04031",
    archivePrefix = "arXiv",
    primaryClass = "astro-ph.CO",
    reportNumber = "FERMILAB-PUB-19-073-AE-E",
    doi = "10.1103/PhysRevD.100.022001",
    journal = "Phys. Rev. D",
    volume = "100",
    number = "2",
    pages = "022001",
    year = "2019"
}

@article{CMS:2022qva,
    author = "Tumasyan, Armen and others",
    collaboration = "CMS",
    title = "{Search for invisible decays of the Higgs boson produced via vector boson fusion in proton-proton collisions at s=13\,\,TeV}",
    eprint = "2201.11585",
    archivePrefix = "arXiv",
    primaryClass = "hep-ex",
    reportNumber = "CMS-HIG-20-003, CERN-EP-2021-273",
    doi = "10.1103/PhysRevD.105.092007",
    journal = "Phys. Rev. D",
    volume = "105",
    number = "9",
    pages = "092007",
    year = "2022"
}

@article{ATLAS:2022yvh,
    author = "Aad, Georges and others",
    collaboration = "ATLAS",
    title = "{Search for invisible Higgs-boson decays in events with vector-boson fusion signatures using 139 fb$^{-1}$ of proton-proton data recorded by the ATLAS experiment}",
    eprint = "2202.07953",
    archivePrefix = "arXiv",
    primaryClass = "hep-ex",
    reportNumber = "CERN-EP-2021-258",
    doi = "10.1007/JHEP08(2022)104",
    journal = "JHEP",
    volume = "08",
    pages = "104",
    year = "2022"
}

@article{Cao:2009yy,
    author = "Cao, Qing-Hong and Ma, Ernest and Shaughnessy, Gabe",
    title = "{Dark Matter: The Leptonic Connection}",
    eprint = "0901.1334",
    archivePrefix = "arXiv",
    primaryClass = "hep-ph",
    doi = "10.1016/j.physletb.2009.02.015",
    journal = "Phys. Lett. B",
    volume = "673",
    pages = "152--155",
    year = "2009"
}

@article{Ibarra:2009bm,
    author = "Ibarra, Alejandro and Ringwald, Andreas and Tran, David and Weniger, Christoph",
    title = "{Cosmic Rays from Leptophilic Dark Matter Decay via Kinetic Mixing}",
    eprint = "0903.3625",
    archivePrefix = "arXiv",
    primaryClass = "hep-ph",
    reportNumber = "DESY-09-039, TUM-HEP-716-09",
    doi = "10.1088/1475-7516/2009/08/017",
    journal = "JCAP",
    volume = "08",
    pages = "017",
    year = "2009"
}

@article{Schmidt:2012yg,
    author = "Schmidt, Daniel and Schwetz, Thomas and Toma, Takashi",
    title = "{Direct Detection of Leptophilic Dark Matter in a Model with Radiative Neutrino Masses}",
    eprint = "1201.0906",
    archivePrefix = "arXiv",
    primaryClass = "hep-ph",
    reportNumber = "KANAZAWA-11-14",
    doi = "10.1103/PhysRevD.85.073009",
    journal = "Phys. Rev. D",
    volume = "85",
    pages = "073009",
    year = "2012"
}

@article{Agrawal:2014ufa,
    author = "Agrawal, Prateek and Chacko, Zackaria and Verhaaren, Christopher B.",
    title = "{Leptophilic Dark Matter and the Anomalous Magnetic Moment of the Muon}",
    eprint = "1402.7369",
    archivePrefix = "arXiv",
    primaryClass = "hep-ph",
    reportNumber = "FERMILAB-PUB-14-016-T",
    doi = "10.1007/JHEP08(2014)147",
    journal = "JHEP",
    volume = "08",
    pages = "147",
    year = "2014"
}

@article{Kopp:2014tsa,
    author = "Kopp, Joachim and Michaels, Lisa and Smirnov, Juri",
    title = "{Loopy Constraints on Leptophilic Dark Matter and Internal Bremsstrahlung}",
    eprint = "1401.6457",
    archivePrefix = "arXiv",
    primaryClass = "hep-ph",
    doi = "10.1088/1475-7516/2014/04/022",
    journal = "JCAP",
    volume = "04",
    pages = "022",
    year = "2014"
}

@article{Baltz:2002we,
    author = "Baltz, E. A. and Bergstrom, L.",
    title = "{Detection of leptonic dark matter}",
    eprint = "hep-ph/0211325",
    archivePrefix = "arXiv",
    reportNumber = "NSF-ITP-02-165",
    doi = "10.1103/PhysRevD.67.043516",
    journal = "Phys. Rev. D",
    volume = "67",
    pages = "043516",
    year = "2003"
}

@article{Chen:2008dh,
    author = "Chen, Chuan-Ren and Takahashi, Fuminobu",
    title = "{Cosmic rays from Leptonic Dark Matter}",
    eprint = "0810.4110",
    archivePrefix = "arXiv",
    primaryClass = "hep-ph",
    reportNumber = "IPMU-08-0071",
    doi = "10.1088/1475-7516/2009/02/004",
    journal = "JCAP",
    volume = "02",
    pages = "004",
    year = "2009"
}

@article{Ko:2010at,
    author = "Ko, Pyungwon and Omura, Yuji",
    title = "{Supersymmetric U(1)B x U(1)L model with leptophilic and leptophobic cold dark matters}",
    eprint = "1012.4679",
    archivePrefix = "arXiv",
    primaryClass = "hep-ph",
    reportNumber = "KIAS-PREPRINT-P10043",
    doi = "10.1016/j.physletb.2011.06.009",
    journal = "Phys. Lett. B",
    volume = "701",
    pages = "363--366",
    year = "2011"
}

@article{Chao:2010mp,
    author = "Chao, Wei",
    title = "{Pure Leptonic Gauge Symmetry, Neutrino Masses and Dark Matter}",
    eprint = "1005.1024",
    archivePrefix = "arXiv",
    primaryClass = "hep-ph",
    doi = "10.1016/j.physletb.2010.10.056",
    journal = "Phys. Lett. B",
    volume = "695",
    pages = "157--161",
    year = "2011"
}

@article{Das:2013jca,
    author = "Das, Moumita and Mohanty, Subhendra",
    title = "{Leptophilic dark matter in gauged $L_{\mu}-L_{\tau}$ extension of MSSM}",
    eprint = "1306.4505",
    archivePrefix = "arXiv",
    primaryClass = "hep-ph",
    doi = "10.1103/PhysRevD.89.025004",
    journal = "Phys. Rev. D",
    volume = "89",
    number = "2",
    pages = "025004",
    year = "2014"
}

@article{Schwaller:2013hqa,
    author = "Schwaller, Pedro and Tait, Tim M. P. and Vega-Morales, Roberto",
    title = "{Dark Matter and Vectorlike Leptons from Gauged Lepton Number}",
    eprint = "1305.1108",
    archivePrefix = "arXiv",
    primaryClass = "hep-ph",
    reportNumber = "FERMILAB-PUB-13-127-T, NUHEP-TH-13-2, UCI-HEP-TR-2013-07, ANL-HEP-PR-13-23",
    doi = "10.1103/PhysRevD.88.035001",
    journal = "Phys. Rev. D",
    volume = "88",
    number = "3",
    pages = "035001",
    year = "2013"
}

@article{Bai:2014osa,
    author = "Bai, Yang and Berger, Joshua",
    title = "{Lepton Portal Dark Matter}",
    eprint = "1402.6696",
    archivePrefix = "arXiv",
    primaryClass = "hep-ph",
    reportNumber = "SLAC-PUB-15912",
    doi = "10.1007/JHEP08(2014)153",
    journal = "JHEP",
    volume = "08",
    pages = "153",
    year = "2014"
}

@article{Bi:2009uj,
    author = "Bi, Xiao-Jun and He, Xiao-Gang and Yuan, Qiang",
    title = "{Parameters in a class of leptophilic models from PAMELA, ATIC and FERMI}",
    eprint = "0903.0122",
    archivePrefix = "arXiv",
    primaryClass = "hep-ph",
    doi = "10.1016/j.physletb.2009.06.009",
    journal = "Phys. Lett. B",
    volume = "678",
    pages = "168--173",
    year = "2009"
}

@article{Kopp:2009et,
    author = "Kopp, Joachim and Niro, Viviana and Schwetz, Thomas and Zupan, Jure",
    title = "{DAMA/LIBRA and leptonically interacting Dark Matter}",
    eprint = "0907.3159",
    archivePrefix = "arXiv",
    primaryClass = "hep-ph",
    reportNumber = "CERN-PH-TH-2009-116",
    doi = "10.1103/PhysRevD.80.083502",
    journal = "Phys. Rev. D",
    volume = "80",
    pages = "083502",
    year = "2009"
}

@article{Fox:2008kb,
    author = "Fox, Patrick J. and Poppitz, Erich",
    title = "{Leptophilic Dark Matter}",
    eprint = "0811.0399",
    archivePrefix = "arXiv",
    primaryClass = "hep-ph",
    reportNumber = "FERMILAB-PUB-08-505-T",
    doi = "10.1103/PhysRevD.79.083528",
    journal = "Phys. Rev. D",
    volume = "79",
    pages = "083528",
    year = "2009"
}

@article{Bell:2019pyc,
    author = "Bell, Nicole F. and Busoni, Giorgio and Robles, Sandra",
    title = "{Capture of Leptophilic Dark Matter in Neutron Stars}",
    eprint = "1904.09803",
    archivePrefix = "arXiv",
    primaryClass = "hep-ph",
    doi = "10.1088/1475-7516/2019/06/054",
    journal = "JCAP",
    volume = "06",
    pages = "054",
    year = "2019"
}

@article{Bell:2021fye,
    author = "Bell, Nicole F. and Busoni, Giorgio and Ramirez-Quezada, Maura E. and Robles, Sandra and Virgato, Michael",
    title = "{Improved treatment of dark matter capture in white dwarfs}",
    eprint = "2104.14367",
    archivePrefix = "arXiv",
    primaryClass = "hep-ph",
    reportNumber = "IPPP/20/54",
    doi = "10.1088/1475-7516/2021/10/083",
    journal = "JCAP",
    volume = "10",
    pages = "083",
    year = "2021"
}

@article{Alloul:2013bka,
    author = "Alloul, Adam and Christensen, Neil D. and Degrande, C\'eline and Duhr, Claude and Fuks, Benjamin",
    title = "{FeynRules  2.0 - A complete toolbox for tree-level phenomenology}",
    eprint = "1310.1921",
    archivePrefix = "arXiv",
    primaryClass = "hep-ph",
    reportNumber = "CERN-PH-TH-2013-239, MCNET-13-14, IPPP-13-71, DCPT-13-142, PITT-PACC-1308",
    doi = "10.1016/j.cpc.2014.04.012",
    journal = "Comput. Phys. Commun.",
    volume = "185",
    pages = "2250--2300",
    year = "2014"
}

@article{Alwall:2014hca,
      author         = "Alwall, J. and Frederix, R. and Frixione, S. and Hirschi,
                        V. and Maltoni, F. and Mattelaer, O. and Shao, H. -S. and
                        Stelzer, T. and Torrielli, P. and Zaro, M.",
      title          = "{The automated computation of tree-level and
                        next-to-leading order differential cross sections, and
                        their matching to parton shower simulations}",
      journal        = "JHEP",
      volume         = "07",
      year           = "2014",
      pages          = "079",
      doi            = "10.1007/JHEP07(2014)079",
      eprint         = "1405.0301",
      archivePrefix  = "arXiv",
      primaryClass   = "hep-ph",
      reportNumber   = "CERN-PH-TH-2014-064, CP3-14-18, LPN14-066, MCNET-14-09,
                        ZU-TH-14-14",
      SLACcitation   = "%%CITATION = ARXIV:1405.0301;%%"
}

@article{Sjostrand:2006za,
      author         = "Sjostrand, Torbjorn and Mrenna, Stephen and Skands, Peter
                        Z.",
      title          = "{PYTHIA 6.4 Physics and Manual}",
      journal        = "JHEP",
      volume         = "05",
      year           = "2006",
      pages          = "026",
      doi            = "10.1088/1126-6708/2006/05/026",
      eprint         = "hep-ph/0603175",
      archivePrefix  = "arXiv",
      primaryClass   = "hep-ph",
      reportNumber   = "FERMILAB-PUB-06-052-CD-T, LU-TP-06-13",
      SLACcitation   = "%%CITATION = HEP-PH/0603175;%%"
}

@article{Sjostrand:2007gs,
      author         = "Sjostrand, Torbjorn and Mrenna, Stephen and Skands, Peter
                        Z.",
      title          = "{A Brief Introduction to PYTHIA 8.1}",
      journal        = "Comput. Phys. Commun.",
      volume         = "178",
      year           = "2008",
      pages          = "852-867",
      doi            = "10.1016/j.cpc.2008.01.036",
      eprint         = "0710.3820",
      archivePrefix  = "arXiv",
      primaryClass   = "hep-ph",
      reportNumber   = "CERN-LCGAPP-2007-04, LU-TP-07-28,
                        FERMILAB-PUB-07-512-CD-T",
      SLACcitation   = "%%CITATION = ARXIV:0710.3820;%%"
}

@article{deFavereau:2013fsa,
    author = "de Favereau, J. and Delaere, C. and Demin, P. and Giammanco, A. and Lema\^\i{}tre, V. and Mertens, A. and Selvaggi, M.",
    collaboration = "DELPHES 3",
    title = "{DELPHES 3, A modular framework for fast simulation of a generic collider experiment}",
    eprint = "1307.6346",
    archivePrefix = "arXiv",
    primaryClass = "hep-ex",
    doi = "10.1007/JHEP02(2014)057",
    journal = "JHEP",
    volume = "02",
    pages = "057",
    year = "2014"
}

@article{Dasgupta:2020dik,
    author = "Dasgupta, Basudeb and Gupta, Aritra and Ray, Anupam",
    title = "{Dark matter capture in celestial objects: light mediators, self-interactions, and complementarity with direct detection}",
    eprint = "2006.10773",
    archivePrefix = "arXiv",
    primaryClass = "hep-ph",
    reportNumber = "TIFR/TH/20-18, ULB-TH/20-07",
    doi = "10.1088/1475-7516/2020/10/023",
    journal = "JCAP",
    volume = "10",
    pages = "023",
    year = "2020"
}

@article{Fan:2010gt,
    author = "Fan, JiJi and Reece, Matthew and Wang, Lian-Tao",
    title = "{Non-relativistic effective theory of dark matter direct detection}",
    eprint = "1008.1591",
    archivePrefix = "arXiv",
    primaryClass = "hep-ph",
    doi = "10.1088/1475-7516/2010/11/042",
    journal = "JCAP",
    volume = "11",
    pages = "042",
    year = "2010"
}

@article{1985ApJ296679P,
       author = {{Press}, W.~H. and {Spergel}, D.~N.},
        title = "{Capture by the sun of a galactic population of weakly interacting, massive particles}",
      journal = "APJ",
       year = 1985,
        month = sep,
       volume = {296},
          doi = {10.1086/163485}
}

@article{1987ApJ321560G,
       author = {{Gould}, Andrew},
        title = "{Weakly Interacting Massive Particle Distribution in and Evaporation from the Sun}",
      journal = "APJ",
      year = 1987,
        month = oct,
       volume = {321},
        pages = {560},
          doi = {10.1086/165652}
}

@article{1987ApJ321571G,
       author = {{Gould}, Andrew},
        title = "{Resonant Enhancements in Weakly Interacting Massive Particle Capture by the Earth}",
      journal = "APJ",
      year = 1987,
        month = oct,
       volume = {321},
        pages = {571},
          doi = {10.1086/165653}
}

@article{Bertone:2007ae,
    author = "Bertone, Gianfranco and Fairbairn, Malcolm",
    title = "{Compact Stars as Dark Matter Probes}",
    eprint = "0709.1485",
    archivePrefix = "arXiv",
    primaryClass = "astro-ph",
    reportNumber = "CERN-PH-TH-2007-158",
    doi = "10.1103/PhysRevD.77.043515",
    journal = "Phys. Rev. D",
    volume = "77",
    pages = "043515",
    year = "2008"
}

@article{Bramante:2017xlb,
    author = "Bramante, Joseph and Delgado, Antonio and Martin, Adam",
    title = "{Multiscatter stellar capture of dark matter}",
    eprint = "1703.04043",
    archivePrefix = "arXiv",
    primaryClass = "hep-ph",
    doi = "10.1103/PhysRevD.96.063002",
    journal = "Phys. Rev. D",
    volume = "96",
    number = "6",
    pages = "063002",
    year = "2017"
}

@article{Dasgupta:2019juq,
    author = "Dasgupta, Basudeb and Gupta, Aritra and Ray, Anupam",
    title = "{Dark matter capture in celestial objects: Improved treatment of multiple scattering and updated constraints from white dwarfs}",
    eprint = "1906.04204",
    archivePrefix = "arXiv",
    primaryClass = "hep-ph",
    reportNumber = "TIFR/TH/19-20",
    doi = "10.1088/1475-7516/2019/08/018",
    journal = "JCAP",
    volume = "08",
    pages = "018",
    year = "2019"
}

@article{Kouvaris:2007ay,
    author = "Kouvaris, Chris",
    title = "{WIMP Annihilation and Cooling of Neutron Stars}",
    eprint = "0708.2362",
    archivePrefix = "arXiv",
    primaryClass = "astro-ph",
    doi = "10.1103/PhysRevD.77.023006",
    journal = "Phys. Rev. D",
    volume = "77",
    pages = "023006",
    year = "2008"
}

@article{Kouvaris:2010vv,
    author = "Kouvaris, Chris and Tinyakov, Peter",
    title = "{Can Neutron stars constrain Dark Matter?}",
    eprint = "1004.0586",
    archivePrefix = "arXiv",
    primaryClass = "astro-ph.GA",
    doi = "10.1103/PhysRevD.82.063531",
    journal = "Phys. Rev. D",
    volume = "82",
    pages = "063531",
    year = "2010"
}

@article{Bell:2018pkk,
    author = "Bell, Nicole F. and Busoni, Giorgio and Robles, Sandra",
    title = "{Heating up Neutron Stars with Inelastic Dark Matter}",
    eprint = "1807.02840",
    archivePrefix = "arXiv",
    primaryClass = "hep-ph",
    doi = "10.1088/1475-7516/2018/09/018",
    journal = "JCAP",
    volume = "09",
    pages = "018",
    year = "2018"
}

@article{Chen:2018ohx,
    author = "Chen, Chian-Shu and Lin, Yen-Hsun",
    title = "{Reheating neutron stars with the annihilation of self-interacting dark matter}",
    eprint = "1804.03409",
    archivePrefix = "arXiv",
    primaryClass = "hep-ph",
    doi = "10.1007/JHEP08(2018)069",
    journal = "JHEP",
    volume = "08",
    pages = "069",
    year = "2018"
}

@article{Acevedo:2019agu,
    author = "Acevedo, Javier F. and Bramante, Joseph and Leane, Rebecca K. and Raj, Nirmal",
    title = "{Warming Nuclear Pasta with Dark Matter: Kinetic and Annihilation Heating of Neutron Star Crusts}",
    eprint = "1911.06334",
    archivePrefix = "arXiv",
    primaryClass = "hep-ph",
    reportNumber = "MIT-CTP/5152",
    doi = "10.1088/1475-7516/2020/03/038",
    journal = "JCAP",
    volume = "03",
    pages = "038",
    year = "2020"
}

@article{deLavallaz:2010wp,
    author = "de Lavallaz, Arnaud and Fairbairn, Malcolm",
    title = "{Neutron Stars as Dark Matter Probes}",
    eprint = "1004.0629",
    archivePrefix = "arXiv",
    primaryClass = "astro-ph.GA",
    doi = "10.1103/PhysRevD.81.123521",
    journal = "Phys. Rev. D",
    volume = "81",
    pages = "123521",
    year = "2010"
}

@article{Graesser:2011wi,
    author = "Graesser, Michael L. and Shoemaker, Ian M. and Vecchi, Luca",
    title = "{Asymmetric WIMP dark matter}",
    eprint = "1103.2771",
    archivePrefix = "arXiv",
    primaryClass = "hep-ph",
    reportNumber = "LA-UR-11-00565",
    doi = "10.1007/JHEP10(2011)110",
    journal = "JHEP",
    volume = "10",
    pages = "110",
    year = "2011"
}

@article{Beltran:2010ww,
    author = "Beltran, Maria and Hooper, Dan and Kolb, Edward W. and Krusberg, Zosia A. C. and Tait, Tim M. P.",
    title = "{Maverick dark matter at colliders}",
    eprint = "1002.4137",
    archivePrefix = "arXiv",
    primaryClass = "hep-ph",
    reportNumber = "FERMILAB-PUB-10-092-A",
    doi = "10.1007/JHEP09(2010)037",
    journal = "JHEP",
    volume = "09",
    pages = "037",
    year = "2010"
}

@article{Goodman:2010ku,
    author = "Goodman, Jessica and Ibe, Masahiro and Rajaraman, Arvind and Shepherd, William and Tait, Tim M. P. and Yu, Hai-Bo",
    title = "{Constraints on Dark Matter from Colliders}",
    eprint = "1008.1783",
    archivePrefix = "arXiv",
    primaryClass = "hep-ph",
    reportNumber = "UCI-HEP-TR-2010-15",
    doi = "10.1103/PhysRevD.82.116010",
    journal = "Phys. Rev. D",
    volume = "82",
    pages = "116010",
    year = "2010"
}

@article{Busoni:2013lha,
    author = "Busoni, Giorgio and De Simone, Andrea and Morgante, Enrico and Riotto, Antonio",
    title = "{On the Validity of the Effective Field Theory for Dark Matter Searches at the LHC}",
    eprint = "1307.2253",
    archivePrefix = "arXiv",
    primaryClass = "hep-ph",
    reportNumber = "CERN-PH-TH-2013-151, SISSA-29-2013-FISI",
    doi = "10.1016/j.physletb.2013.11.069",
    journal = "Phys. Lett. B",
    volume = "728",
    pages = "412--421",
    year = "2014"
}

@article{Shoemaker:2011vi,
    author = "Shoemaker, Ian M. and Vecchi, Luca",
    title = "{Unitarity and Monojet Bounds on Models for DAMA, CoGeNT, and CRESST-II}",
    eprint = "1112.5457",
    archivePrefix = "arXiv",
    primaryClass = "hep-ph",
    reportNumber = "LA-UR-11-12298",
    doi = "10.1103/PhysRevD.86.015023",
    journal = "Phys. Rev. D",
    volume = "86",
    pages = "015023",
    year = "2012"
}

@article{Contino:2016jqw,
    author = "Contino, Roberto and Falkowski, Adam and Goertz, Florian and Grojean, Christophe and Riva, Francesco",
    title = "{On the Validity of the Effective Field Theory Approach to SM Precision Tests}",
    eprint = "1604.06444",
    archivePrefix = "arXiv",
    primaryClass = "hep-ph",
    reportNumber = "DESY-16-067, CERN-TH-2016-082, LPT-Orsay-16-32",
    doi = "10.1007/JHEP07(2016)144",
    journal = "JHEP",
    volume = "07",
    pages = "144",
    year = "2016"
}

@article{CMS:2018yxg,
    author = "Sirunyan, Albert M and others",
    collaboration = "CMS",
    title = "{Search for an $L_{\mu}-L_{\tau}$ gauge boson using Z$\to4\mu$ events in proton-proton collisions at $\sqrt{s} =$ 13 TeV}",
    eprint = "1808.03684",
    archivePrefix = "arXiv",
    primaryClass = "hep-ex",
    reportNumber = "CMS-EXO-18-008, CERN-EP-2018-208",
    doi = "10.1016/j.physletb.2019.01.072",
    journal = "Phys. Lett. B",
    volume = "792",
    pages = "345--368",
    year = "2019"
}

@article{Altmannshofer:2016jzy,
    author = "Altmannshofer, Wolfgang and Gori, Stefania and Profumo, Stefano and Queiroz, Farinaldo S.",
    title = "{Explaining dark matter and B decay anomalies with an $L_\mu - L_\tau$ model}",
    eprint = "1609.04026",
    archivePrefix = "arXiv",
    primaryClass = "hep-ph",
    reportNumber = "MITP-16-065",
    doi = "10.1007/JHEP12(2016)106",
    journal = "JHEP",
    volume = "12",
    pages = "106",
    year = "2016"
}

@article{PandaX:2022xqx,
    author = "Li, Shuaijie and others",
    collaboration = "PandaX",
    title = "{Search for Light Dark Matter with Ionization Signals in the PandaX-4T Experiment}",
    eprint = "2212.10067",
    archivePrefix = "arXiv",
    primaryClass = "hep-ex",
    doi = "10.1103/PhysRevLett.130.261001",
    journal = "Phys. Rev. Lett.",
    volume = "130",
    number = "26",
    pages = "261001",
    year = "2023"
}

@article{PandaX:2023xgl,
    author = "Huang, Di and others",
    collaboration = "PandaX",
    title = "{Search for Dark-Matter\textendash{}Nucleon Interactions with a Dark Mediator in PandaX-4T}",
    eprint = "2308.01540",
    archivePrefix = "arXiv",
    primaryClass = "hep-ex",
    doi = "10.1103/PhysRevLett.131.191002",
    journal = "Phys. Rev. Lett.",
    volume = "131",
    number = "19",
    pages = "191002",
    year = "2023"
}

@article{LZ:2024zvo,
    author = "Aalbers, J. and others",
    collaboration = "LZ",
    title = "{Dark Matter Search Results from 4.2 Tonne-Years of Exposure of the LUX-ZEPLIN (LZ) Experiment}",
    eprint = "2410.17036",
    archivePrefix = "arXiv",
    primaryClass = "hep-ex",
    reportNumber = "FERMILAB-PUB-24-0796-V",
    month = "10",
    year = "2024"
}

@article{InternationalMuonCollider:2024jyv,
    author = "Accettura, Carlotta and others",
    collaboration = "International Muon Collider",
    title = "{Interim report for the International Muon Collider Collaboration (IMCC)}",
    eprint = "2407.12450",
    archivePrefix = "arXiv",
    primaryClass = "physics.acc-ph",
    reportNumber = "CERN-2024-002",
    doi = "10.23731/CYRM-2024-002",
    journal = "CERN Yellow Rep. Monogr.",
    volume = "2/2024",
    pages = "176",
    year = "2024"
}

@article{CCFR:1991lpl,
    author = "Mishra, S. R. and others",
    collaboration = "CCFR",
    title = "{Neutrino Tridents and W Z Interference}",
    reportNumber = "NEVIS-1437, FERMILAB-PUB-91-390",
    doi = "10.1103/PhysRevLett.66.3117",
    journal = "Phys. Rev. Lett.",
    volume = "66",
    pages = "3117--3120",
    year = "1991"
}

@article{Roy:2024ear,
    author = "Roy, Arnab and Dasgupta, Basudeb and Guchait, Monoranjan",
    title = "{Constraining Asymmetric Dark Matter using colliders and direct detection}",
    eprint = "2402.17265",
    archivePrefix = "arXiv",
    primaryClass = "hep-ph",
    doi = "10.1007/JHEP08(2024)095",
    journal = "JHEP",
    volume = "08",
    pages = "095",
    year = "2024"
}

@article{Hapitas:2021ilr,
    author = "Hapitas, Timothy and Tuckler, Douglas and Zhang, Yue",
    title = "{General kinetic mixing in gauged U(1)L{\ensuremath{\mu}}-L{\ensuremath{\tau}} model for muon g-2 and dark matter}",
    eprint = "2108.12440",
    archivePrefix = "arXiv",
    primaryClass = "hep-ph",
    doi = "10.1103/PhysRevD.105.016014",
    journal = "Phys. Rev. D",
    volume = "105",
    number = "1",
    pages = "016014",
    year = "2022"
}

@article{Falkowski:2011xh,
    author = "Falkowski, Adam and Ruderman, Joshua T. and Volansky, Tomer",
    title = "{Asymmetric Dark Matter from Leptogenesis}",
    eprint = "1101.4936",
    archivePrefix = "arXiv",
    primaryClass = "hep-ph",
    reportNumber = "LPT-ORSAY-11-09",
    doi = "10.1007/JHEP05(2011)106",
    journal = "JHEP",
    volume = "05",
    pages = "106",
    year = "2011"
}

@article{Delahaye:2019omf,
    author = "Delahaye, Jean Pierre and Diemoz, Marcella and Long, Ken and Mansouli{\'e}, Bruno and Pastrone, Nadia and Rivkin, Lenny and Schulte, Daniel and Skrinsky, Alexander and Wulzer, Andrea",
    title = "{Muon Colliders}",
    eprint = "1901.06150",
    archivePrefix = "arXiv",
    primaryClass = "physics.acc-ph",
    month = "1",
    year = "2019"
}

@article{Long:2020wfp,
    author = "Long, K. and Lucchesi, D. and Palmer, M. and Pastrone, N. and Schulte, D. and Shiltsev, V.",
    title = "{Muon colliders to expand frontiers of particle physics}",
    eprint = "2007.15684",
    archivePrefix = "arXiv",
    primaryClass = "physics.acc-ph",
    reportNumber = "FERMILAB-PUB-20-366-AD-APC",
    doi = "10.1038/s41567-020-01130-x",
    journal = "Nature Phys.",
    volume = "17",
    number = "3",
    pages = "289--292",
    year = "2021"
}

@article{MuonCollider:2022xlm,
    author = "de Blas, Jorge and others",
    collaboration = "Muon Collider",
    title = "{The physics case of a 3 TeV muon collider stage}",
    eprint = "2203.07261",
    archivePrefix = "arXiv",
    primaryClass = "hep-ph",
    reportNumber = "FERMILAB-CONF-22-317-AD-ND-PPD-SCD-TD",
    month = "3",
    year = "2022"
}

@article{Accettura:2023ked,
    author = "Accettura, Carlotta and others",
    title = "{Towards a muon collider}",
    eprint = "2303.08533",
    archivePrefix = "arXiv",
    primaryClass = "physics.acc-ph",
    reportNumber = "FERMILAB-PUB-23-123-AD-PPD-T",
    doi = "10.1140/epjc/s10052-023-11889-x",
    journal = "Eur. Phys. J. C",
    volume = "83",
    number = "9",
    pages = "864",
    year = "2023",
    note = "[Erratum: Eur.Phys.J.C 84, 36 (2024)]"
}

@article{Han:2020uak,
    author = "Han, Tao and Liu, Zhen and Wang, Lian-Tao and Wang, Xing",
    title = "{WIMPs at High Energy Muon Colliders}",
    eprint = "2009.11287",
    archivePrefix = "arXiv",
    primaryClass = "hep-ph",
    reportNumber = "PITT-PACC 2009",
    doi = "10.1103/PhysRevD.103.075004",
    journal = "Phys. Rev. D",
    volume = "103",
    number = "7",
    pages = "075004",
    year = "2021"
}

@article{Capdevilla:2024bwt,
    author = "Capdevilla, Rodolfo and Meloni, Federico and Zurita, Jose",
    title = "{Discovering Electroweak Interacting Dark Matter at Muon Colliders Using Soft Tracks}",
    eprint = "2405.08858",
    archivePrefix = "arXiv",
    primaryClass = "hep-ph",
    reportNumber = "FERMILAB-PUB-23-0832-T, DESY-24-069",
    doi = "10.1103/PhysRevLett.134.181802",
    journal = "Phys. Rev. Lett.",
    volume = "134",
    number = "18",
    pages = "181802",
    year = "2025"
}

@article{Casarsa:2021rud,
    author = "Casarsa, Massimo and Fabbrichesi, Marco and Gabrielli, Emidio",
    title = "{Monochromatic single photon events at the muon collider}",
    eprint = "2111.13220",
    archivePrefix = "arXiv",
    primaryClass = "hep-ph",
    doi = "10.1103/PhysRevD.105.075008",
    journal = "Phys. Rev. D",
    volume = "105",
    number = "7",
    pages = "075008",
    year = "2022"
}

@article{Jana:2023ogd,
    author = "Jana, Sudip and Klett, Sophie",
    title = "{Muonic force and nonstandard neutrino interactions at muon colliders}",
    eprint = "2308.07375",
    archivePrefix = "arXiv",
    primaryClass = "hep-ph",
    doi = "10.1103/PhysRevD.110.095011",
    journal = "Phys. Rev. D",
    volume = "110",
    number = "9",
    pages = "095011",
    year = "2024"
}

@article{Jueid:2023zxx,
    author = "Jueid, Adil and Nasri, Salah",
    title = "{Lepton portal dark matter at muon colliders: Total rates and generic features for phenomenologically viable scenarios}",
    eprint = "2301.12524",
    archivePrefix = "arXiv",
    primaryClass = "hep-ph",
    reportNumber = "CTPU-PTC-2023-02",
    doi = "10.1103/PhysRevD.107.115027",
    journal = "Phys. Rev. D",
    volume = "107",
    number = "11",
    pages = "115027",
    year = "2023"
}

@article{Asadi:2023csb,
    author = "Asadi, Pouya and Radick, Aria and Yu, Tien-Tien",
    title = "{Interplay of freeze-in and freeze-out: Lepton-flavored dark matter and muon colliders}",
    eprint = "2312.03826",
    archivePrefix = "arXiv",
    primaryClass = "hep-ph",
    doi = "10.1103/PhysRevD.110.035022",
    journal = "Phys. Rev. D",
    volume = "110",
    number = "3",
    pages = "035022",
    year = "2024"
}

@article{Asadi:2024jiy,
    author = "Asadi, Pouya and Homiller, Samuel and Radick, Aria and Yu, Tien-Tien",
    title = "{Fermion-portal dark matter at a high-energy muon collider}",
    eprint = "2412.14235",
    archivePrefix = "arXiv",
    primaryClass = "hep-ph",
    doi = "10.1103/p7p8-wqqb",
    journal = "Phys. Rev. D",
    volume = "112",
    number = "5",
    pages = "055038",
    year = "2025"
}

@article{Liu:2022byu,
    author = "Liu, Jiao and Han, Zhi-Long and Jin, Yi and Li, Honglei",
    title = "{Unraveling the Scotogenic model at muon collider}",
    eprint = "2207.07382",
    archivePrefix = "arXiv",
    primaryClass = "hep-ph",
    doi = "10.1007/JHEP12(2022)057",
    journal = "JHEP",
    volume = "12",
    pages = "057",
    year = "2022"
}

@article{Huang:2021nkl,
    author = "Huang, Guo-yuan and Queiroz, Farinaldo S. and Rodejohann, Werner",
    title = "{Gauged $L^{}_{\mu}{-}L^{}_{\tau}$ at a muon collider}",
    eprint = "2101.04956",
    archivePrefix = "arXiv",
    primaryClass = "hep-ph",
    doi = "10.1103/PhysRevD.103.095005",
    journal = "Phys. Rev. D",
    volume = "103",
    number = "9",
    pages = "095005",
    year = "2021"
}

@article{Feng:2012jn,
    author = "Feng, Wan-Zhe and Nath, Pran and Peim, Gregory",
    title = "{Cosmic Coincidence and Asymmetric Dark Matter in a Stueckelberg Extension}",
    eprint = "1204.5752",
    archivePrefix = "arXiv",
    primaryClass = "hep-ph",
    doi = "10.1103/PhysRevD.85.115016",
    journal = "Phys. Rev. D",
    volume = "85",
    pages = "115016",
    year = "2012"
}

@article{Dasgupta:2023zrh,
    author = "Dasgupta, Arnab and Dev, P. S. Bhupal and Han, Tao and Padhan, Rojalin and Wang, Si and Xie, Keping",
    title = "{Searching for heavy leptophilic Z': from lepton colliders to gravitational waves}",
    eprint = "2308.12804",
    archivePrefix = "arXiv",
    primaryClass = "hep-ph",
    reportNumber = "PITT-PACC-2317, IP/BBSR/2023-09",
    doi = "10.1007/JHEP12(2023)011",
    journal = "JHEP",
    volume = "12",
    pages = "011",
    year = "2023"
}

@article{Bell:2025acg,
    author = "Bell, Nicole F. and Busoni, Giorgio and Ghosh, Avirup",
    title = "{Using neutron stars to probe dark matter charged under a L{\ensuremath{\mu}}-L{\ensuremath{\tau}} symmetry}",
    eprint = "2505.06506",
    archivePrefix = "arXiv",
    primaryClass = "hep-ph",
    doi = "10.1088/1475-7516/2025/10/060",
    journal = "JCAP",
    volume = "10",
    pages = "060",
    year = "2025"
}

@article{Garani:2019fpa,
    author = "Garani, Raghuveer and Heeck, Julian",
    title = "{Dark matter interactions with muons in neutron stars}",
    eprint = "1906.10145",
    archivePrefix = "arXiv",
    primaryClass = "hep-ph",
    reportNumber = "ULB-TH/19-05, UCI-TR-2019-17",
    doi = "10.1103/PhysRevD.100.035039",
    journal = "Phys. Rev. D",
    volume = "100",
    number = "3",
    pages = "035039",
    year = "2019"
}

@article{Bell:2020lmm,
    author = "Bell, Nicole F. and Busoni, Giorgio and Robles, Sandra and Virgato, Michael",
    title = "{Improved Treatment of Dark Matter Capture in Neutron Stars II: Leptonic Targets}",
    eprint = "2010.13257",
    archivePrefix = "arXiv",
    primaryClass = "hep-ph",
    doi = "10.1088/1475-7516/2021/03/086",
    journal = "JCAP",
    volume = "03",
    pages = "086",
    year = "2021"
}

@article{Belfkir:2023vpo,
    author = "Belfkir, Mohamed and Jueid, Adil and Nasri, Salah",
    title = "{Boosting dark matter searches at muon colliders with machine learning: The mono-Higgs channel as a case study}",
    eprint = "2309.11241",
    archivePrefix = "arXiv",
    primaryClass = "hep-ph",
    reportNumber = "CTPU-PTC-23-37",
    doi = "10.1093/ptep/ptad144",
    journal = "PTEP",
    volume = "2023",
    number = "12",
    pages = "123B03",
    year = "2023"
}

@article{Dawson:2025dmi,
    author = "Dawson, Sally and Roy, Arnab and Valencia, German",
    title = "{Semi-visible higgs decay as a probe for new invisible particles}",
    eprint = "2511.09778",
    archivePrefix = "arXiv",
    primaryClass = "hep-ph",
    month = "11",
    year = "2025"
}

@article{Wang:2025cth,
    author = "Wang, X. G. and Thomas, A. W.",
    title = "{A Viable New Model for Dark Matter}",
    eprint = "2510.24114",
    archivePrefix = "arXiv",
    primaryClass = "hep-ph",
    reportNumber = "ADP-25-34/T1296",
    month = "10",
    year = "2025"
}

@article{Guchait:2020wqn,
    author = "Guchait, Monoranjan and Roy, Arnab",
    title = "{Light Singlino Dark Matter at the LHC}",
    eprint = "2005.05190",
    archivePrefix = "arXiv",
    primaryClass = "hep-ph",
    doi = "10.1103/PhysRevD.102.075023",
    journal = "Phys. Rev. D",
    volume = "102",
    number = "7",
    pages = "075023",
    year = "2020"
}

@article{Heeck:2011wj,
    author = "Heeck, Julian and Rodejohann, Werner",
    title = "{Gauged  $L_\mu  -  L_\tau$  Symmetry  at  the  Electroweak  Scale}",
    eprint = "1107.5238",
    archivePrefix = "arXiv",
    primaryClass = "hep-ph",
    doi = "10.1103/PhysRevD.84.075007",
    journal = "Phys. Rev. D",
    volume = "84",
    pages = "075007",
    year = "2011"
}

@article{Muong-2:2025xyk,
    author = "Aguillard, D. P. and others",
    collaboration = "Muon g-2",
    title = "{Measurement of the Positive Muon Anomalous Magnetic Moment to 127~ppb}",
    eprint = "2506.03069",
    archivePrefix = "arXiv",
    primaryClass = "hep-ex",
    reportNumber = "FERMILAB-PUB-25-0364-PPD",
    doi = "10.1103/7clf-sm2v",
    journal = "Phys. Rev. Lett.",
    volume = "135",
    number = "10",
    pages = "101802",
    year = "2025"
}

@article{Aliberti:2025beg,
    author = "Aliberti, R. and others",
    title = "{The anomalous magnetic moment of the muon in the Standard Model: an update}",
    eprint = "2505.21476",
    archivePrefix = "arXiv",
    primaryClass = "hep-ph",
    reportNumber = "CERN-TH-2025-101, FERMILAB-PUB-25-0344-T, INT-PUB-25-015, IPARCOS-UCM-25-029, KEK Preprint 2025-22, LTH 1403, MITP-25-037, UWThPh 2025-15, UWThPh
  2025-15, ZU-TH 37/25, IPARCOS-UCM-25-029",
    doi = "10.1016/j.physrep.2025.08.002",
    journal = "Phys. Rept.",
    volume = "1143",
    pages = "1--158",
    year = "2025"
}

@article{De:2025hay,
  author        = "De, Bibhabasu",
  title         = "{Exploring the null results in the direct detection experiments, $(g-2)_{\ell}$ and neutrino mass in an extended $\textrm{U}(1)_{L_{\mu}-L_{\tau}}$ model constrained through the $Z\to \ell^{+}\ell^{-}$ decays}",
  eprint        = "2509.10939",
  archivePrefix = "arXiv",
  primaryClass  = "hep-ph",
  doi           = "10.1007/JHEP12(2025)075",
  journal       = "JHEP",
  volume        = "12",
  pages         = "075",
  year          = "2025"
}

@article{Chen:2025ewx,
    author = "Chen, Wanyun and Li, Haoqi and Lu, Chih-Ting and Wang, Qiulei",
    title = "{Exploring muonphilic dark matter with the $Z_2$-even mediator at muon colliders}",
    eprint = "2511.21290",
    archivePrefix = "arXiv",
    primaryClass = "hep-ph",
    month = "11",
    year = "2025"
}
\end{document}